\documentstyle[12pt,aasms,psfig, epsf]{article}
\begin{document}

\title{On the Spatial Distribution of Hard X-Rays from Solar Flare Loops}
\author{Vah\'{e} Petrosian\altaffilmark{1} and Timothy Q. 
Donaghy\altaffilmark{2}}
\affil{Center for Space Science and Astrophysics\\
Stanford University\\
Stanford, CA 94305-4060}
\date{today}

\altaffiltext{1}{Also Department of Physics and Applied Physics}
\altaffiltext{2}{Also Department of Physics}

\newcommand{\beq}{\begin{equation}}
\newcommand{\eeq}{\end{equation}}
\newcommand{\lsim}{\mbox{$\stackrel{<}{\scriptstyle\sim}$}}
\newcommand{\gsim}{\mbox{$\stackrel{>}{\scriptstyle\sim}$}}
\newcommand{\D}[2]{\makebox{$\displaystyle\frac{\partial{#1}}{\partial{#2}}$}}
\newcommand{\DD}[2]{\makebox{$\displaystyle\frac{\partial^2{#1}}{\partial{#2}^2}
$}}

\begin{abstract}

The aim of this paper is to investigate  the spatial structure of the impulsive 
phase hard X-ray emission from solar flares. This work is motivated by the 
YOHKOH and the 
forthcoming HESSI observations.
Summarizing past results, it is shown that the transport effects can account for 
the observations by inhomogeneous loops where there is a strong field 
convergence and/or density enhancement at the top of  the flaring loop. 
Scattering by plasma turbulence at the acceleration site or pancake type pitch 
angle distribution of the accelerated electrons can also give rise to enhanced 
emission at the loop tops. These could be a natural consequence of acceleration 
by plasma waves.
This paper considers a general case of stochastic scattering and  
acceleration that leads to an isotropic  pitch angle distribution and  an 
enhanced emission from the loop tops or the acceleration site.

Following the formalism developed in earlier papers the strength and the 
spectrum of the radiation expected from the acceleration site and the foot 
points are evaluated and their dependence on the parameters describing the 
acceleration process and the flare plasma are determined. The theoretical ratio 
of these two intensities and relative values of their spectral indices are 
compared with the YOHKOH observations, demonstrating that the above mentioned 
parameters can be constrained with such observations. It is shown that future 
high spatial and spectral resolution observations, for example those expected 
from HESSI, can begin to distinguish between different models and constrain 
their parameters.

{\it Subject Headings:} acceleration of 
particles--plasmas--waves--Sun:flares--Sun:X-rays, gamma rays
\end{abstract}

\section{Introduction}

It is well established that electrons and protons are accelerated to 
relativistic energies during the impulsive phase of solar flares.
However, the source of energy and the acceleration mechanism are not well 
understood.
In the standard thick target model, it is assumed that the acceleration arises 
in the reconnection region somewhere above a flaring loop.
Unfortunately, these loops are observed and delineated not by the direct 
radiation from the 
accelerated particles (hard X-rays, gamma-rays and microwave radiation) but by 
the thermal emissions (soft X-rays, optical, etc.) from plasmas heated and 
evaporated by the dissipation in the 
lower corona and the chromosphere of the energy of the accelerated particles.
High spatial resolution observations of the direct radiation  to fully 
map the flares during the impulsive phase are more difficult.
Nonetheless, there exist some concrete results which, for the most part, agree 
with the above picture.

There have been several high spatial resolution observations at the microwave 
region by OVRO, VLA and BIMA radio telescopes.
However, interpretation of the spatial distribution of the microwave emissions 
is complicated by the fact that it depends not only on the characteristics of 
the accelerated electrons, but also strongly on the strength and geometry of the 
magnetic field.
In the case of symmetric loops, the appearance of circularly polarized radiation 
from the loops (Marsh \& Hurford, 1980) agrees with the standard model of 
accelerated electrons injected at the top of the loop (Petrosian, 1982).
More recent observations have revealed foot point emission primarily from one 
end of the 
loop, indicating the presence of strong asymmetry such that most of the 
radiation 
arises from the foot point with the higher magnetic field, emphasizing the 
dependence on the field strength and its variations.
We shall not deal with microwave radiation in this paper.

Here we are concerned with the spatially resolved hard X-ray observations.
First resolved images of hard X-ray impulsive emissions were provided by the SMM 
and HINOTORI space crafts, showing that in majority of flares most of the 
emission 
arises from two foot point regions as expected in the standard thick target 
model.
A more reliable confirmation of this picture has come from the YOHKOH satellite, 
which  has detected impulsive hard X-ray 
emission from an area near the top of the loop, as well as the foot points, from 
several flares (Masuda et al. 1995).
In general, this loop top emission is weaker and is more readily detectable in 
limb flares.
Masuda (1994) finds loop top emission in six of ten such flares, indicating that 
this characteristic is common and may be the rule rather than the exception.

We believe that the next significant advance of our knowledge of the impulsive 
phase processes will come from more observations at higher energies and with 
higher spatial resolutions, such as those expected from HESSI.
Understanding of this new data will require a complete analysis of spatial 
distribution of hard X-ray emission from a flaring loop.

This paper deals with this subject in general and with the relative strength and 
 spectral characteristics of the loop top vs. foot point emissions observed by 
YOHKOH in particular.
The modeling of this phenomenon is discussed in \S 2, where we first review past 
relevant theoretical results and then introduce a model where plasma turbulence 
plays a crucial role.
In \S 3 we describe this model and in \S 4 we compare results from 
it with the YOHKOH observations.
A brief summary and discussion is given in \S 5.

\section{Spatial Distribution of Hard X-ray Emissions}

For interpretation of these observations we need the spatial distribution of 
hard X-ray emission during the impulsive phase by electrons as they traverse the 
loop from the acceleration site to the foot points.
There have been several studies of this subject with some recent ones 
specifically designed to replicate the YOHKOH results (e.g. Masuda et al. 1994; 
Wheatland \& Melrose 1994; Fletcher 1995; Holman 1996; and Fletcher \& Martens 
1998).
In this section we will review these and earlier studies, most of which assume 
injection of some high energy electrons with an assumed energy and pitch angle 
distribution at the top of a symmetric closed loop.

\subsection{Model Parameters}

Several factors enter in the determination of the relative emissivity of the 
hard X-rays along the loop at different energies.  
The effects of almost all of these factors were first investigated in the Ph.D. 
thesis 
of J. Leach (1984) and in several related publications (Leach \& Petrosian 1981 
and 1983).
It was assumed that the acceleration process injects electrons at the top of a 
loop, with the distribution $f(E, \alpha, s=0)$, as a function of energy $E$ (in 
units of the electron rest mass energy $m_{e} c^{2}$) and pitch angle $\alpha$.

In this case, the most important factors are the parameters describing the 
distributions of the accelerated electrons (e.g. spectral index $\delta$, the 
beaming direction $\alpha _{b}$ and width $\alpha _{0}$), and  the variation of 
plasma density $n(s)$ and magnetic field $B(s)$ along the loop.
Two important parameters related to the characteristics of the plasma are $d \ln 
B / d \ln N$, where \( N(s) = \int_{0}^{s} n(x) dx \) is the column depth 
measured along the loop from the injection point at $s=0$, and $N_{tr}$  the 
column depth to the transition region. One may replace the first parameter by an 
average convergence rate; e.g. the ratio $B(N_{tr})/B(0)$.
At higher (gamma ray) energies and for microwave emission, the absolute value of 
the magnetic field is also important (see e.g. McTiernan \& Petrosian 1990a and 
1990b, and Lu \& Petrosian 1989).
In this section we are dealing with photon energies $k < 100$ keV where this 
will not be important.
We now review the effects of these parameters in isolation and together.

\subsection{Uniform Loops}

\begin{figure}[htbp]
\leavevmode
\centerline{
\psfig{file=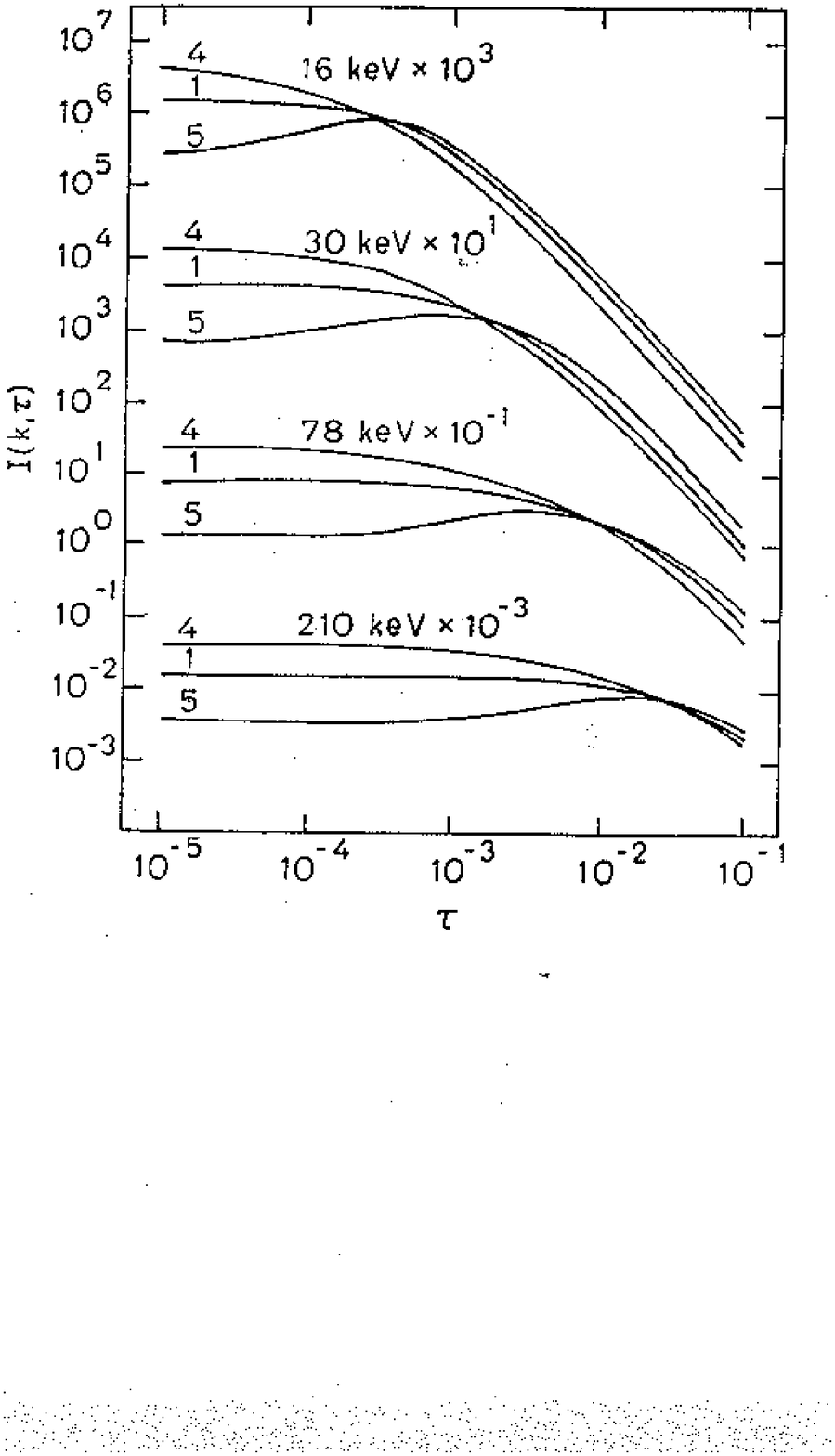,width=0.45\textwidth,height=0.75\textwidth}\vspace{-6.0 cm}
\psfig{file=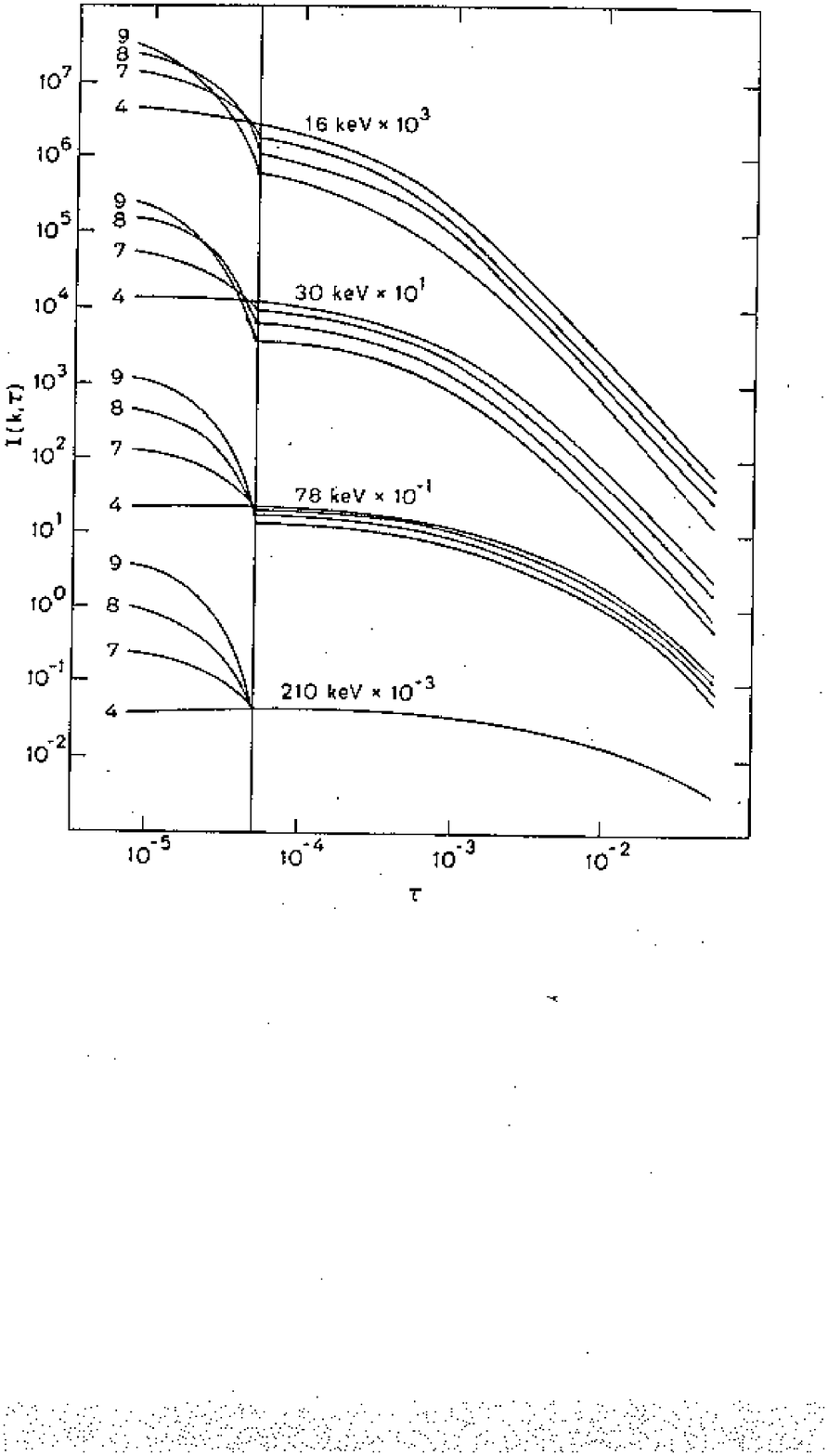,width=0.45\textwidth,height=0.75\textwidth}\vspace{-6.0 cm}}
\vspace{-6.0 cm}
\caption{These Figures taken from the thesis of J. Leach (1984) shows the 
variation of normalized intensity with dimensionless column depth $\tau = 
N/N_{0}$, where $N_0=(4\pi r_0^2 {\rm ln}\Lambda)^{-1}= 5\times 10^{22} {\rm 
cm}^{-2}$. The curves are labeled by the values of the photon energies and for 
clarity are shifted upward by the indicated amounts. {\it Left Panel} 
demonstrates the effects of the degree of beaming of the accelerated electrons; 
curve marked 4, 1 and 5 are for three models with values of the pitch angle 
distribution widths $\alpha_0^2= \infty, 0.4$  and 0.04, respectively. The 
magnetic field is uniform and the spectral index $\delta=5$.
{\it right Panel} shows the effects of magnetic field convergence for models 
with $\delta=5$ and isotropic pitch angle 
distribution; the field 
convergence rate is chosen to give ratios of magnetic field strength at the 
transition 
region to that at the top of the loop $B(N_{tr})/B(0) = 1$, 2, 5 and 25 for 
curves with labels 4, 7, 8 and 9, respectively.
}
\label{leach}
\end{figure}

For loops with constant field strengths, $B(s) = B_{0}$, the variation of the 
X-ray intensity along the loop becomes simple when expressed in terms of the 
column depth $N$.
Figure \ref{leach}, from Leach (1984), shows the variation with dimensionless 
column 
depth, $\tau = N/N_{0}$, of the normalized intensity of  hard X-rays 
($\int_{0}^{\infty} I( \tau, k) d\tau = 1$), for various photon energies $k$ 
and for three models. 
Here $N_{0} = (4 \pi r_{0}^2 \ln \Lambda )^{-1}=5\times 10^{22} $cm$^{-2}$, 
where
$r_{0}$ is the classical radius of the electron,  $\ln \Lambda \simeq 20$ is 
the Coulomb logarithm. The injected electron distribution is assumed to obey
$f(E, \alpha, s=0) \propto E^{-\delta} 
\exp(-(\alpha-\alpha_b)^{2} / \alpha_{0}^{2})$, with $\delta = 5$. The left 
panel of this figure shows the effects of the degree of the beaming of the 
accelerated electrons for uniform magnetic field loops; 
$\alpha_{b}=0$ and $\alpha_{0}^{2}=0.04$, $0.4$ and $\infty$ (i.e.
 an isotropic pitch angle distribution), for curves labeled 5, 1 and 4, 
respectively.
Leach (1984) finds that within 10 to 30\%, the curves for the intermediate model
$(\alpha_0^2=0.4)$ and for $\delta=3$, $4$ and $5$ can be described by
\begin{equation}
\label{diffJ}
I(N, k)dN = (\frac{\delta}{2}-1) \left( 1 + 
\frac{N}{N(\alpha_{0}^{2}, k)} \right) ^{-\delta / 2} 
\frac{dN}{N(\alpha_{0}^{2}, k)},
\end{equation}
where $N(\alpha_{0}^{2}, k)$ is the column depth below which most of the photons 
with energy $k$ originate; the intensity is constant for smaller column depths 
but decreases rapidly for column depths exceeding this value.  

$N(\alpha_{0}^{2},k)$ depends weekly on the pitch angle distribution of the 
injected electrons.
For the intermediate model, Leach finds that $N(0.4, k) \simeq N_{0} 
k^{2}/(k+1)$.
For the isotropic model $N(\infty, k) \simeq 0.37 N_{0} k^{2}/(k+1)$ which as 
expected is smaller so that the intensity begins to drop earlier.
The reverse is true for the model with strong beaming along the field lines; the 
critical column depth is about a factor of 4 higher. Here and in what follows 
the photon energy $k$ is in units of $m_ec^2$.
The above expression fails at very low column depths for the last model because 
the 
Coulomb collisions gradually scatter the strongly beamed electrons into higher 
pitch 
angles, giving rise to some small, initial increases of intensity with 
increasing $N$ or $\tau$.
This  dependence on both the spectral index and pitch angle distribution 
was further investigated by McTiernan \& Petrosian (1990b) and Petrosian, 
McTiernan \& Marschh\"auser (1994) at higher photon energies, and more recently 
by Fletcher (1996) in an analysis involving variation 
of the average height of the hard X-ray emission with photon energy and model 
parameters.

To convert these distributions to ones in terms of distance $s$ along 
the loop, we need to specify the density variation $n(s)$.
For example, for a truly uniform loop $n(s) = n_{0}$ and $N =n_0s$ so that the 
above behavior reflects the variations with $s$ as well.
In a real loop, because of the high coronal temperature, the density scale 
height (or more exactly $ds / d \ln n$) is expected to be much larger than the 
length of the loop, $L\sim 10^9 {\rm cm}$, so that the constancy of the density 
in the coronal 
portion is reasonable.
However, this condition fails completely below the transition region where the 
density first jumps abruptly and then increases exponentially with a much 
smaller ($\sim 10^{8}$ cm) scale height. For photon energies where the critical 
cplumn depth 
$N(\alpha_{0}^{2},k) > N_{tr}$, the column depth to the transition region, 
we expect a nearly uniform or very slowly decreasing flux throughout the 
loop followed by an abrupt increase at the foot points.
On the other hand for lower photon energies, or for $N(\alpha_{0}^{2}, k) < 
N_{tr}$ most of the emission will come from the loop rather than the foot 
points.
Integration of equation (\ref{diffJ}) gives the ratio of emission throughout the 
coronal portion of the loop to that from the transition region and below it 
(i.e. foot points);
\begin{equation} \label{LP/FP}
\frac{J_{Loop}}{J_{FP}} = -1 + \left(1+\frac{N_{tr}}{N(\alpha_{0}^{2},k)} 
\right) 
^{\delta / 2 -1},\, \,  \, \, \delta > 2.
\end{equation}
Since for most flares, and in particular for the YOHKOH limb flares of interest 
here, this ratio is less than one, the quantity $N_{tr}$ must be less than 
$N(\alpha_{0}^{2},k)$.
This also means that whatever emission there is from the loop will be very 
nearly uniform and not concentrated towards the top of the loop.

One possible way to obtain substantial radiation from the loop top but not from 
the sides of the loop, is to inject electrons with a ``pancake'' distribution 
$(\alpha _{b} = \pi / 2$ and $\alpha _{0} \ll 1)$.
Leach (1984) did not consider such models. The consequences of this 
type of distribution were considered for studies of the directivity of the high 
energy emissions (hard X-rays and gamma rays) by Dermer \& Ramaty (1986) and 
McTiernam \& 
Petrosian (1991), and in the investigation of the average height of lower energy 
X-ray emission by Fletcher (1996).
We shall return to this possibility below.

\subsection{Non-Uniform Loops}

The simple spatial variation described above can be  modified considerably if 
the 
plasma density and/or the magnetic field vary strongly along the loop.
For example, if the density decreases from the loop top to the foot point in the 
coronal part of the loop, then the thin target emission from this portion being 
proportional to the ambient density will 
mimic this variation; equation (\ref{diffJ}) will still be valid, but the
spatial distribution will be $I(s, k) = I(N, k)n(s)$.
Such a model was proposed by Wheatland and Melrose (1995) for the YOHKOH limb 
flares.
It is difficult to see how large density gradients can be maintained in a $T 
\geq 10^{6} K$ plasma unless the density enhancement is a transient phenomena.

Enhancement of loop top emission can be produced more naturally in loops with 
converging field geometry.
The field convergence increases the pitch angle of the particles as they travel 
down the loop, and if strong enough, can reflect them back up the loop before 
they reach the transition region.
This increases the density of the  accelerated electrons near the top of the 
loop 
giving rise to a corresponding enhancement of the bremsstrahlung emission.
Such models were also investigated by Leach (1984), showing significant 
increases of emission from loop tops (right panel Figure \ref{leach}), and by 
Fletcher (1996) 
showing increase in the average height of the X-ray emission.

Finally, trapping of electrons near the loop top can also come about by an 
enhanced pitch angle scattering.
The primary reason that the hard X-ray bremsstrahlung emission is described by 
the smooth and monotonic expression of equation (\ref{diffJ}) is due to the fact 
that for Coulomb collisions (the dominant interaction process for 
nonrelativistic energies) the mean free paths for the pitch angle scattering 
and energy loss
are comparable and approximately equal to $\lambda 
_{\mathrm{Coul}} = -vE/\dot{E}_{\mathrm{Coul}}$ where $v=\beta 
c$ is the electron velocity and 
$\dot{E}_{\rm Coul}=-4\pi r_0^2 n c{\rm ln}\Lambda/\beta$ is the energy 
loss rate (in units of $m_ec^2 $).
However, if there exists another scattering agent with $\lambda _{\mathrm{scat}} 
\ll \lambda _{\mathrm{Coul}}$ then electrons undergo many pitch angle 
scatterings before they lose a substantial amount of their energy.
Furthermore, if $\lambda _{\mathrm{scat}} \ll L$, the size of the region near 
the top of the 
loop, the electrons will be trapped for a time $T_{\rm esc} \simeq \frac{L}{v} 
\times  \frac{L}{\lambda _{\mathrm{scat}}} \gg \frac{L}{v}$, achieve an 
isotropic pitch angle distribution ($\alpha_0 \rightarrow \infty$), and give 
rise to 
enhanced emission from the loop top.
The ratio of the loop top emission compared to emission from other parts of the 
coronal portion of the loop (which do not contain such a scattering agent but 
are of comparable length) will be approximately $J_{LT} / J_{Loop} \simeq 
L/\lambda_{\mathrm{scat}}$, in which case we may detect emission only from the 
top of the loop and the high density foot points, but not from the legs of the 
loop.

In summary, in order to have the bulk of the emission to come from the foot 
points we need $N_{tr} \ll N(\alpha_{0}^{2},k) \simeq N_0 k^{2}/(k+1)$, which is 
equivalent to having $\lambda _{\mathrm{Coul}} \gg L$ in the coronal parts of 
the loop.
And in order to enhance the emission near the loop top without doing the same 
from the 
rest of the loop we require $\lambda _{\mathrm{scat}} \ll L$.
A possible scattering agent which could satisfy this condition is plasma
turbulence.
If this is the case, however, one must also include the possibility of 
acceleration of the particles by the same plasma  turbulence.

\section{Stochastic Acceleration}

Stochastic acceleration of particles by plasma waves or turbulence, a second 
order Fermi acceleration processes, is one of the leading candidates for 
production of nonthermal particles in solar flares.
Their role in accelerating protons and ions have been investigated in some 
simplified models by Ramaty and collaborators (see e.g. Ramaty 1979 and Ramaty 
\& Murphy 1987).
Based on some theoretical arguments and on new higher spectral and spatial 
resolution data (such as that from the events under consideration here)  we have 
argued that stochastic acceleration by plasma turbulence is the most promising 
process for acceleration of flare electrons (Petrosian 1994 and 1996).
Similar arguments have been put forth by Schlickeiser and collaborators 
(Schlickeiser 1989; Bech et al. 1990), and Miller and collaborators (Miller 
1991; Miller, LaRosa \& Moore 1996; Miller \& Reames 1996).
In a combined acceleration, transport and radiation model we have compared 
theoretical spectra with observations (Petrosian et al.
1994; Park, Petrosian \& Schwartz 
1997, hereafter {\bf PPS}) and have shown that some of the observed X-ray and 
gamma 
ray 
spectral features, in particular the deviations from a simple power law, can be 
most naturally explained with such models.
PPS explore various acceleration models and include the 
effects due to Coulomb collisions and synchrotron losses and determine model 
parameters for several flares observed by GRS 
on SMM, and BATSE and EGRET on C-GRO.
Here we use the same formalism to determine the spatial distribution of hard 
X-rays.
We first review the relevant aspects of this model and then compare its 
predictions with the YOHKOH observations.

\subsection{Model Description}

The exact solution of the combined acceleration, transport and radiation 
processes 
requires solution of the time dependent coupled kinetic equations for waves and 
particles.
This is beyond the scope of this work and is not warranted for comparison with 
the existing data.
We therefore make the following simplifying and somewhat justifiable 
assumptions.
We shall return to the possible breakdown of these assumptions in the summary 
section.

{\bf a)} We assume that the flare energizing process, presumably magnetic 
reconnection, produces plasma turbulence throughout the impulsive phase at a 
rate faster than the damping time of the turbulence, so that the observed 
variation of the impulsive phase emissions is due to modulation of the 
energizing process.
This assumption decouples the kinetic equation of waves from that of the 
electrons.
We assume that the turbulence is confined to a region of size  $L$ near the top 
of 
the flaring loop.

{\bf b)} The requirement $\lambda _{\rm {scat}} \ll L$ indicates that the plasma
electrons will undergo repeated pitch angle scattering in the acceleration 
region so that a nearly isotropic pitch angle distribution is established.
Therefore we need to contend only with diffusion in momentum (or energy) due to 
the turbulence and energy losses by other processes.
The kinetic equation for diffusion in the acceleration region is (for further 
details see Park \& Petrosian  1995 and 1996, and PPS)
\begin{equation} \label{KEQ}
\frac{\partial f}{\partial t} = \frac{\partial ^2}{\partial E^2} [D(E) f] - 
\frac{\partial}{\partial E}[(A(E) - |\dot{E}_{L}|)f] \nonumber \\ 
-\frac{f}{T_{\rm esc} (E)} + Q(E).
\end{equation}
Here $D$ and $A$ are the diffusion and systematic acceleration coefficients due 
to the stochastic process which are obtained from the standard Fokker-Planck 
equation 
(Chandrasekhar 1943), and 
\begin{equation}\label{loss}
\dot{E}_{L} = \dot{E}_{\rm Coul} + \dot{E}_{\rm synch} = - 4\pi 
r_{0}^{2} c n \ln \Lambda / \beta - 4r_0^2B^{2}\beta^{2} 
\gamma^{2}/9m_ec^2 
\end{equation}
describes the  Coulomb  and synchrotron energy loss rates (in units of $m_ec^2 
$).
For the source function, $Q(E)$, we use a Maxwellian distribution with 
temperature, $kT = 1.5$ keV.

{\bf c)} We will be comparing model results with ``maps'' integrated over 
a time which is much longer than all other time scales such as the 
traversal time $\tau _{tr} = \frac{L}{c \beta} = 0.03 {\rm s}(\frac{L}{10^{9} 
cm})\frac{1}{\beta}$, the escape time $T_{\rm esc} \simeq (\frac{L}{\lambda 
_{\rm scat}})\tau _{tr}$, or the acceleration time $\tau _{ac} \simeq 
\frac{D(E)}{E^2} \simeq \frac{A(E)}{E}$.
Therefore, we can ignore the time dependence and use the steady state solution 
$(\frac{\partial f}{\partial t} = 0)$.

{\bf d)} Following PPS we use the whistler model but mostly a  more general 
model with 
\beq
\label{coeffs}
D(E)={\cal D} \beta (\gamma \beta)^{q'} ,  \,\,\,
A(E)={\cal D} (q' +2)(\gamma \beta)^{q' -1}, \,\,\,
T_{\rm esc}(E)={\cal T}_{\rm esc}(\gamma \beta)^s/\beta + L/(\beta c \sqrt{2}). 
\eeq
Here we have added some average traverse time as a minimum escape time and have 
used a different notation than PPS; our $q' $ plays the role of the 
parameter $q$ in PPS, while we reserve $q$ for the spectral index of the 
(assumed) power law distribution for the plasma wave vectors (PPS use the symbol 
$\tilde q$ for this latter index). Our notation here agrees with that in Dung \& 
Petrosian (1994) and Pryadko \& Petrosian (1997, 1998 and 1999). As shown in 
these papers, the rate of acceleration is proportional to ${\cal D}$ which in 
turn is proportional to the electron gyrofrequency, $\Omega_e$, and the level of 
turbulence $f_{turb}=8\pi {\cal E}_{tot}/B^2$, where ${\cal E}_{tot}$ is the 
total energy density of turbulence. In the
Whistler model ${\cal D}$ and ${\cal T}_{\rm esc}$ are related and $s=q' = 
q - 2$.  For a more complete analysis one should use the
numerical results given in the last four references which evaluate the total 
contributions from all plasma modes.  The
existing data do not warrant considerations of such details.  Consequently, we
shall leave such details for the interpretation of better data to be obtained by
HESSI.  However, we shall choose the parameters in the above expressions so that
the relevant parameters qualitatively behave like the ones from these numerical
results.

\begin{figure}[htbp]
\leavevmode
\centerline{
\psfig{file=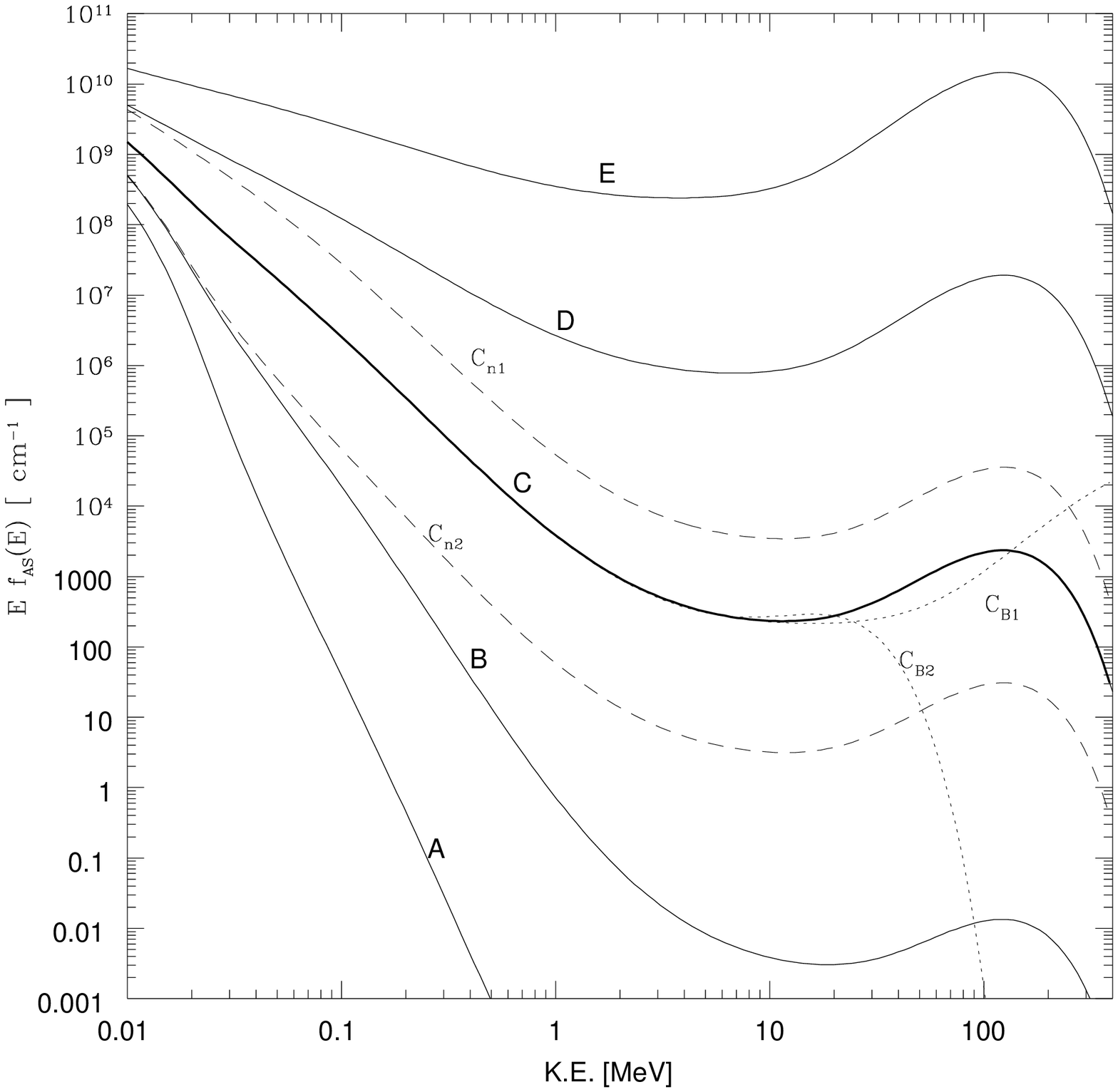,width=0.5\textwidth,height=0.5\textwidth}
\psfig{file=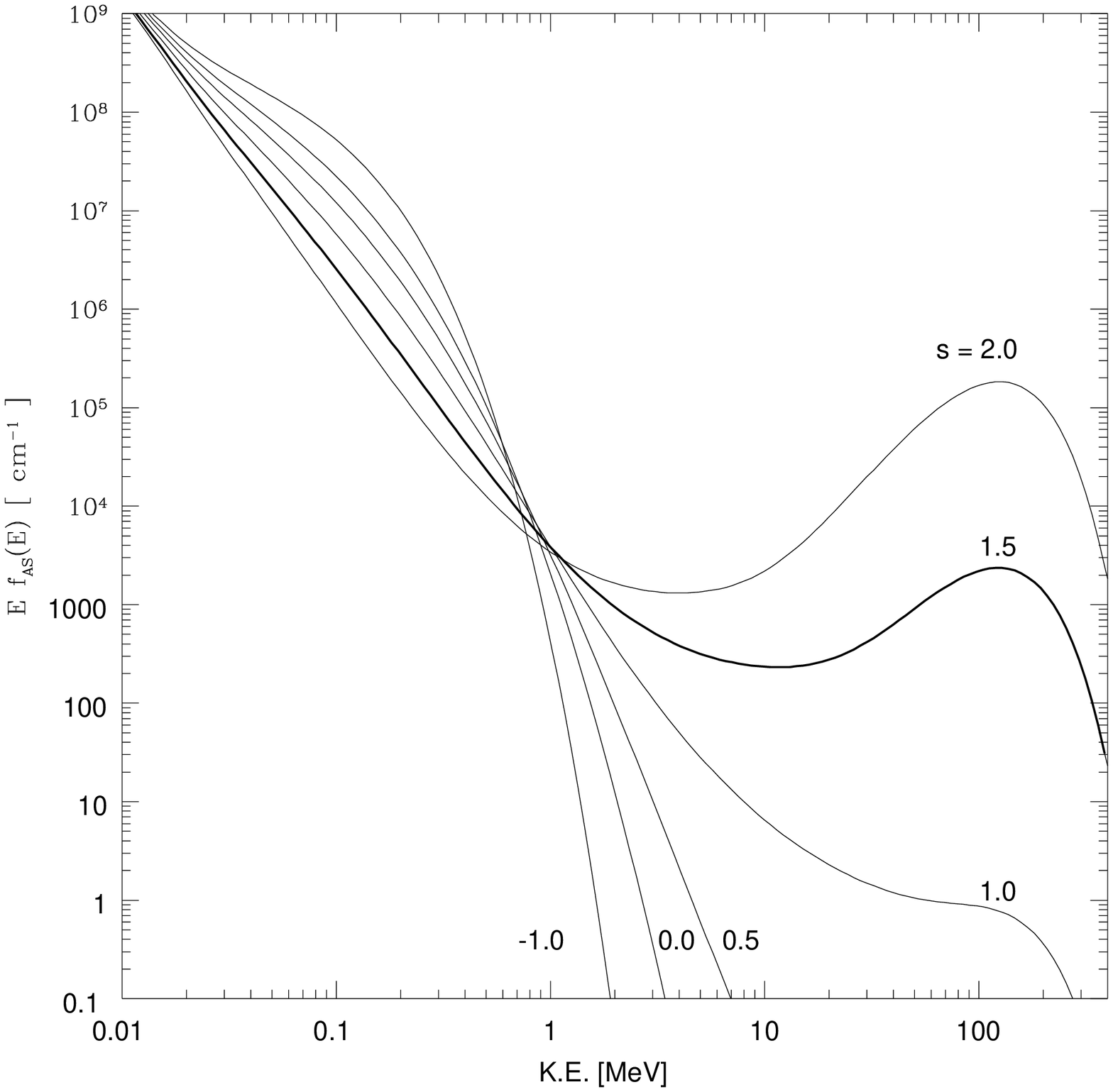,width=0.5\textwidth,height=0.5\textwidth}
}
\caption{Representative  spectra of the accelerated electrons for two different 
kinds of models. For convenience we plot product of energy times spectral 
density. {\it Left Panel:} The solid lines ($A$ to $E$) are for models with 
$\zeta_{\rm 
esc} = 0.0025$, 0.005, 0.01, 0.02 and 
0.04, respectively. We use a constant ratio of  $\zeta_{\rm Coul}/\zeta_{\rm 
esc} 
=167$ or $n{\cal T}_{\rm esc}=10^{10}$ cm$^{-3}$ s, so that the proportionality 
factor in equations (\ref{ratio}) and (\ref{ratio2}) is constant. We use 
$B=300$G or 
$\zeta_{\rm synch} = 420$, $q' =1.7$ and $s=1.5$; these exponent are chosen 
to 
qualitatively mimic the behavior of the Fokker-Planck coefficients for a steep 
spectrum of turbulence ($q=3$) calculated by Pryadko \& Petrosian (1997). The 
two dashed 
lines $C_{n1}$ and $C_{n2}$ are same as model $C$ except the density is lower 
and higher by factor of 2, respectively. The dotted lines labeled $C_{B1}$ and 
$C_{B2}$ are 
for magnetic fields of 100 and 1000 G, respectively. 
The parameters for the various curves can also be found in Table 1.
{\it Right Panel:} Curves are labeled by the value of the exponent $s$ showing 
the effects of the energy dependence of the ratio of the accleration to escape 
times; i.e. the exponent $s+q'$. All other parameters same as model $C$ (heavy 
soild lines) in both panels.
}
\label{particle}
\end{figure}

The spectrum of the accelerated particles depends primarily on the  
three dimensionless parameters

\beq
\label{dimless}
\zeta_{\rm esc}={\cal DT}_{\rm esc},\,\,\,
\zeta_{\rm Coul}={\cal D}/(4\pi r_0^2cn{\rm ln}\Lambda), \,\,\,
\zeta_{\rm synch}=9{\cal D}m_ec^2 /(4r_0^2cB^2),
\eeq
on the exponents $s$ and $q' $, and weakly on the size $L$ (only at very 
low energies).
Figure \ref{particle} shows some representative spectra of the electrons in the 
acceleration
site, $f_{AS}(E)$, obtained from solving equation (\ref{KEQ}).  If we ignore the 
losses ( i.e.  for $\zeta_{\rm
Coul}$ and $\zeta_{\rm synch}$ $\gg 1$ ) the only parameters affecting the
spectrum are $\zeta_{\rm esc}$ and $s+q' $ which determine the amplitude 
and the
energy dependence of the ratio of the escape time to the acceleration time;
$R_{\rm esc}\equiv DT_{\rm esc}/E^2$. 

In the left panel of Figure \ref{particle} we use $q' =1.7, s=1.5$ so
that this ratio decreases with $E$ at nonrelativistic energies but
increases at extreme relativistic values mimicking the steep turbulence spectrum
($q=3$) results of Pryadko \& Petrosian (1997). The solid curves (labeled $A$ to 
$E$) show the the spectra for five values of $\zeta_{\rm esc}=0.0025, 0.05, 0.1, 
0.2$ and 0.4, respectively. Other parameters are kept constant, except we vary 
the density so that $n{\cal T}_{\rm esc} \propto \zeta_{\rm esc}/\zeta_{\rm 
Coul}$ remains constant.  These spectra are
steep throughout for small values of $\zeta_{\rm esc}$, but they flatten  with
increasing $\zeta_{\rm esc}$ and may acquire a positive power law index at 
relativistic energies. All spectra eventually fall off exponentially when the 
synchrotron loss rate becomes comparable to the acceleration rate; i.e. $R_{\rm 
synch}\equiv -D(E)/(E \dot E_{\rm synch}) \rightarrow 1$. Using relativistic 
approximation it can 
be shown that this happens for energies beyond the 
critical energy  
$E_{\rm synch}=\zeta_{\rm synch}^{1/(3-q' )}$.  This effect is evident more 
dramatically from comparison of the two dotted curves labeled $C_{B1}$ and 
$C_{B2}$ with the heavy solid line $C$ which have same parameters except the 
magnetic field is set to 100, 1000 and 300, respectively.
In a similar manner,  the Coulomb losses steepen the spectrum below the critical 
energy $E_{\rm Coul}$ obtained from the equality $R_{\rm Coul}\equiv -D(E)/(E 
\dot{E}_{\rm Coul})=1$.
For nonrelativistic energies 
$E_{\rm Coul}=2^{-(2+q' )/q' } \zeta_{\rm Coul}^{-2/q' }$. This 
effect can be seen by comparing the solid curve $C$ with the dashed lines 
labeled $C_{n1}$ and $C_{n2}$ which are for three different values of density. 
For a complete listing of the parameters used in this and subsequent related 
figures see Table 1. (For further discussion of these effects see  Hamilton \& 
Petrosian 1992 and Bech et al. 1990).

In the right panel of Figure \ref{particle} we show the effect of the variation 
of the energy dependence of the 
ratio $R_{\rm esc}$. Here we vary the exponent $s$  and keep all other 
parameters constant and equal to their values used for the model represented by 
the heavy solid line labeled $C$ in both panels of  Figure \ref{particle}. For 
larger exponent 
the acceleration time becomes shorter than the escape time at high energies 
and gives rise to a larger number of high energy electrons. For lower values of 
this exponent the ratio $R_{\rm esc}$ decreases
monotonicly with energy both at nonrelativistic and extreme relativistic
cases. This behavior is qualitatively similar to what one may obtain for a flat
spectrum of turbulence such as the case $q=5/3$ in Pryadko \& Petrosian (1997).
 In this case the electrons escape too quickly for
a substantial acceleration to high energies.  Consequently, the spectra fall off
steeply at high energies. This will be true  for all values of $\zeta_{\rm esc}$ 
used in Figure \ref{particle}.  However, as
expected the spectra flatten at low energies. The extremely efficient 
acceleration of low energy particles is responsible for this behavior in these 
models. These two figures demonstrate that a rich variety of accelerated 
electron spectra are possible by this mechanism.

\begin{figure}[htbp]
\leavevmode
\centerline{
\psfig{file=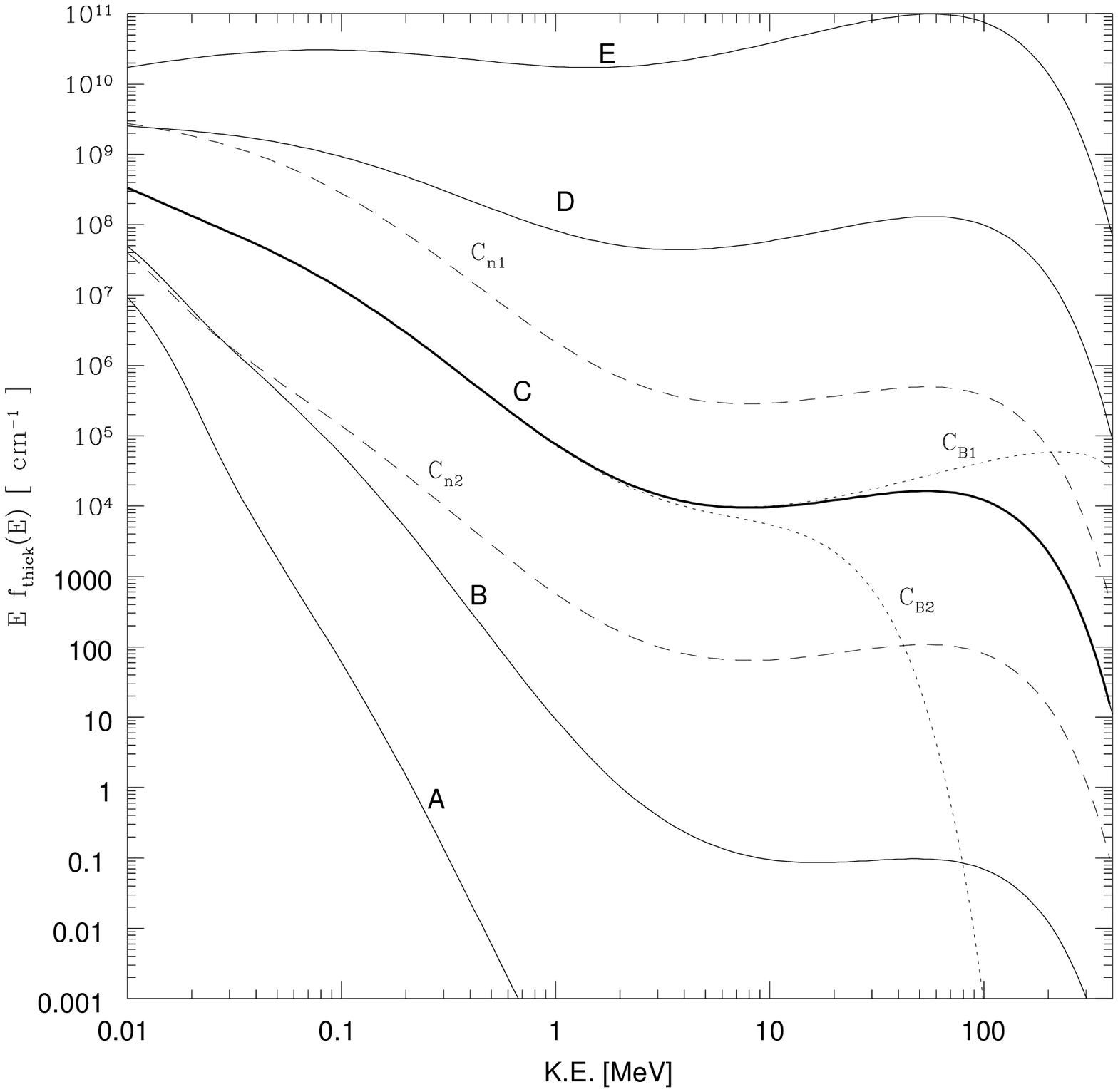,width=0.5\textwidth,height=0.5\textwidth}
\psfig{file=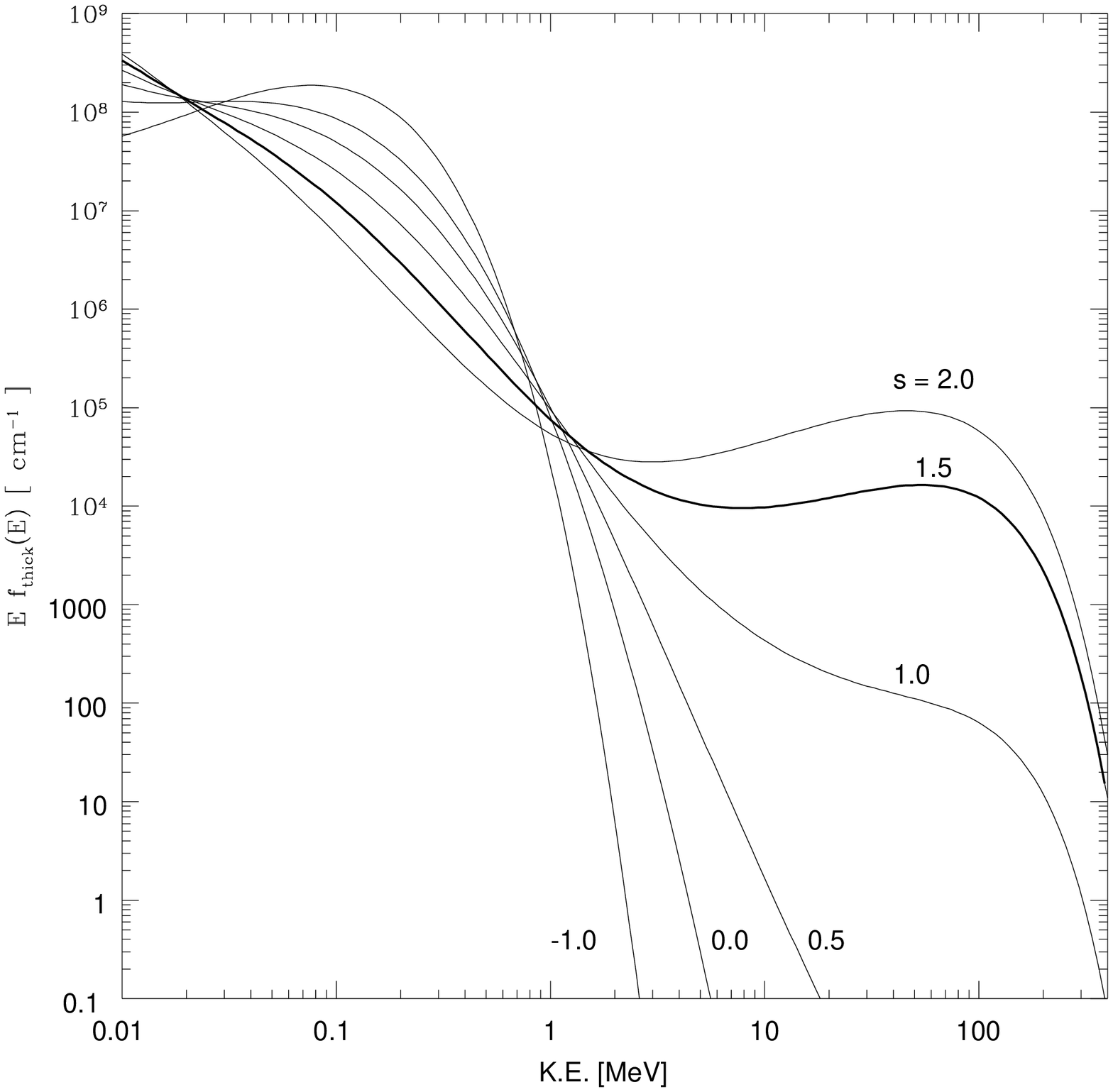,width=0.5\textwidth,height=0.5\textwidth}
}
\caption{The effective electron  spectra at the thick target (foot points) as 
defined by equation (\ref{fthick}) for the  models of Figure \ref{particle} or 
the parameters given in Table 1. In general these spectra are flatter than those 
in Figure \ref{particle}.}
\label{thick}
\end{figure}

\begin{center}
\centerline{Table 1}
\centerline{Model Parameters for left panels of Figures \ref{particle}, 
\ref{thick}, \ref{spectra} and \ref{specratio}}
\begin{tabular}{|l|c|c|c|c|c|c|c|c|} \hline\hline
Label & ${\cal T}_{esc} [s]$ & $n [cm^{-3}]$ & $B [gauss]$ & $\zeta _{esc}$ 
& $\zeta_{Coul}$ & $\zeta_{synch}$ & $E_{Coul}[m_{e}c^{2}]$ & 
$E_{synch}[m_{e}c^{2}]$ \\\hline
A 	 & 0.05 & $2 \times 10^{11}$    & 300  & 0.0025 & 0.417 & 421  & 1.68 & 
104 \\
B 	 & 0.10 & $1 \times 10^{11}$    & 300  & 0.005  & 0.833 & 421  & 0.88 & 
104 \\
C 	 & 0.20 & $5 \times 10^{10}$    & 300  & 0.01   & 1.667 & 421  & 0.26 & 
104 \\
D 	 & 0.40 & $2.5 \times 10^{10}$  & 300  & 0.02   & 3.333 & 421  & 0.058 & 
104 \\
E 	 & 0.80 & $1.25 \times 10^{10}$ & 300  & 0.04   & 6.667 & 421  & 0.025 & 
104 \\\hline
$C_{B1}$ & 0.20 & $5 \times 10^{10}$    & 100  & 0.01   & 1.667 & 3788 & 0.26 & 
565 \\
$C_{B2}$ & 0.20 & $5 \times 10^{10}$    & 1000 & 0.01   & 1.667 & 37.9 & 0.26 & 
16.4 \\\hline
$C_{n1}$ & 0.20 & $2.5 \times 10^{10}$  & 300  & 0.01   & 3.333 & 421  & 0.058 & 
104 \\
$C_{n2}$ & 0.20 & $1 \times 10^{11}$    & 300  & 0.01   & 0.833 & 421  & 0.88 & 
104 \\
\hline\hline
\end{tabular}
\newline
${\cal D} = 0.05[s^{-1}]$,\
$L = 1 \times 10^{9}$ [cm],\
$q' = 1.7$,\
$s = 1.5$,\
$Q_{0} = 1 \times 10^{12}$ [particles/sec]
\end{center}

In what follows we will explore the model parameters represented in the left 
panel of Figure \ref{particle} 
because of its diversity and because the spectral fittings by PPS obtained a 
range of values closer to  this region of the parameter space. 

The escaping electrons will have a nearly isotropic pitch angle distribution in
the downward direction and a spectral flux equal to $F_{\rm esc}=Lf_{AS}/T_{\rm
esc}$.  As we shall see below for the calculation of the thick target 
bremsstrahlung
emission by these electrons at the foot points we shall need the integral of
this flux divided by the loss term. We call this $f_{\rm thick}$:

\beq \label{fthick} f_{\rm thick}= - { 1 \over {\dot E}_L} \int^\infty_E{f_{AS}
\over T_{\rm esc}}dE \eeq 
This spectrum is shown in Figure \ref{thick} for the same
parameters used in Figure \ref{particle}.  As evident there is considerable 
differences
between $f_{AS}$ and $f_{\rm thick}$.  
For simple power laws, i.e. for low values of $\zeta_{\rm esc}$, the power law 
index of spectrum relevant for the foot points 
is expected to be smaller (in absolute value, i.e. flatter) than that of the 
spectrum at the acceleration site by $ 2-s/2$ and $1-s$ for 
nonrelativistic and extreme relativistic limits, respectively. Thus, in 
general, unless the exponent $s$ for
the escape term is greater than 4, we expect the foot point spectrum to be 
flatter than that at the acceleration site. For intermediate 
values of 
$\zeta_{\rm esc}$  both $f_{AS}$ and $f_{\rm thick}$ undergo several changes in 
their spectral indices. Both spectra undergo further variations at low 
and high 
energies where the effects of the Coulomb and synchrotron losses come in.
These differences between the two spectra will produce
similar differences between the emissions from the acceleration site and the
foot points.

\subsection{Bremsstrahlung Emission}

We now evaluate the emission from the acceleration site in 
addition to that from the coronal portion of the loop and the foot points 
discussed earlier.

{\bf The Thin Target} bremsstrahlung photon number spectrum from the 
acceleration site can be 
found from
\begin{equation}
\label{LoopTop}
  J_{AS}(k) = L \int_{k}^{\infty} dE f_{AS}(E) \beta c
    n_{AS} \frac{d\sigma}{dk}(E, k) ,
\end{equation}
where $n_{AS}$ is the background density in the acceleration site, $d\sigma/dk$, 
given by Koch \& Motz (1959, eq. [3BN]),
is the relativistically correct, angle integrated,
bremsstrahlung cross section,
$L$ is the length of the acceleration region at the top of the loop so that $V=A 
\times L$ is the volume of the acceleration region. 
The spectrum $f_{AS}$ is integrated over the cross-sectional area of the loop so 
that the area $A$  only  affects the overall normalization
of the photon spectrum and can be absorbed into the source function $Q$ in 
equation
(\ref{KEQ}). 

\begin{figure}[htbp]
\leavevmode\centering
\psfig{file=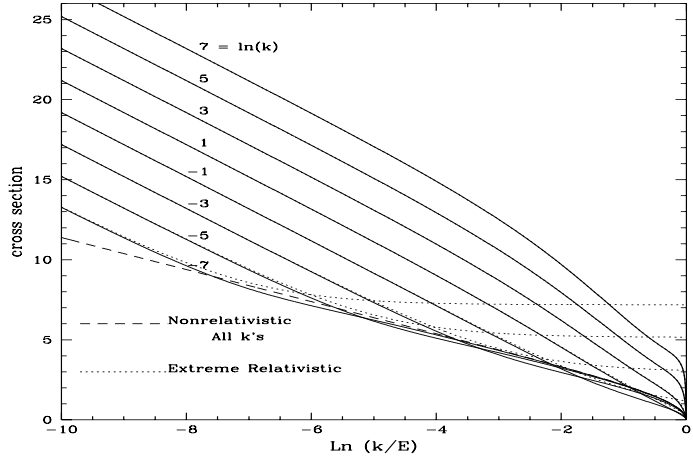,width=0.6\textwidth,height=0.4\textwidth}
\caption{Comparison of the extreme relativistic (dotted lines) and the 
nonrelativistic (dashed lines) approximations for the bremsstrahlung cross 
section given by the right hand side of equation (9) with the more 
accurate formula given in Koch \& Motz (1959, eq. [3BN]) (solid lines).}
\label{cross}
\end{figure}

In the nonrelativistic and extreme relativistic limits the bremsstrahlung cross 
section can be 
approximated as
\begin{equation}
\label{crosssec}
\frac{3 \beta^{2} k}{16 \alpha r_{0}^{2}} \frac{d\sigma}{dk} = 
\left\{ \begin{array}{ll}
	\ln \frac{1+\sqrt{1-x}}{1-\sqrt{1-x}} & \mbox{for $k$, $E \ll 1$,}\\
	\left(\ln\frac{2k}{\sqrt{e}} + \ln\frac{1-x}{x^2} 
\right)(1-x+\frac{3}{4} x^{2}) & \mbox{for $E \gg 1$.}
	\end{array}
\right.
\end{equation}
where $\alpha$ is the fine structure constant and $x=k/E \leq 1$ is the ratio of 
photon to electron energies. Figure \ref{cross} compares
these expressions with the more accurate 3BN formula of Koch and Motz (1959). 
The nonrelativistic approximation is valid only for very low values of both $k$ 
and 
$E$. Its range of validity decreases rapidly with increasing $k$. For 
ln$k=-3$, (photon energy of $\sim 188$ keV) it is within $\leq 10\%$ of the 
exact value for 
electron energies $E<1$. Therefore, some caution is required in 
using this approximation for photon energies $\geq 100$ keV. On the other hand, 
the extreme relativistic approximation is 
not only valid for $k$ and $E \gg 1$ but also at nonrelativistic photon 
energies as long as $E \gg 1$. For $k \sim 1$  it requires $E > 3$  
for a 10\% or better accuracy.

As mentioned above a nearly isotropic (in the downward direction) flux of 
electrons,  
$F_{esc} (E)=Lf_{AS}/T_{\rm esc}$,  will escape the acceleration site and will 
be 
transported along the loop to the foot points, emitting bremsstrahlung radiation 
with the spatial distribution described in the previous section (e.g. eq.[1]).
As long as $N _{tr}$, the column depth from the edge of the acceleration site to 
the transition region, is significantly less than $N(\alpha_{0}^{2}, k) \simeq 
N_{0} k^{2}/(k+1)$, most of this radiation will come from the foot points.
As stated previously, this condition requires that $\lambda _{\mathrm{loss}}$ 
($=\lambda _{\mathrm{Coul}}$ for nonrelativistic electrons) to be much larger 
than 
the length of the leg of the loop (which is of the order of $L$). This means 
that 
the electrons are moving essentially freely downward and their isotropic 
distribution is preserved even if the field converges: The field convergence has 
no significant effect because the flux integrated over the tube cross-section is 
conserved (see e.g. Leach \& Petrosian 1983).
Thus, most of the radiation will come from the foot points in a thick target 
source.

{\bf The Thick Target} bremsstrahlung spectrum of this source is obtained from 
the following well known expression (see e.g. Petrosian 1973)
\begin{equation}
\label{FootPoint}
  J_{FP}(k) = L\int_{k}^{\infty} dE
    f_{\rm thick}\beta c n\frac{d\sigma}{dk}(E, k).
\end{equation}
The density $n$ here and in $\dot E_L$ used in equation (\ref{fthick}) refer to 
the 
background density at the
foot points. Here the density may exceed $10^{12}$ 
cm$^{-3}$ so that for field strengths of $B \simeq 300$G the synchrotron 
losses 
become important only for Lorentz factors $\gamma \geq 10^2$, which is beyond 
even 
the EGRET observations considered in PPS. Consequently, for all relevant 
energies, in particular for 
nonrelativistic electrons under consideration here, we have $\frac{\dot E_L}{n} 
\simeq \frac{\dot E_{coul}}{n} = -4 \pi r_{0}^{2} \ln \Lambda c/\beta $ which 
means that the foot point emission will be independent of the plasma parameters 
$n$ and $B$ (except, of course, for the dependence of $f_{\rm thick}$ on these 
quantities).

\subsection{Loop Top vs. Foot point Emission}

We now determine the intensity  and spectral differences between the
acceleration site and foot point emissions. In what follows we first derive some 
analytic expressions using power law approximations for the electron spectra and 
ignoring the  slowly (logarithmic) varying part of the cross-section (right hand 
side 
of
equation [9]). We then present some numerically calculated spectra 
using the exact expressions for the electron spectra and the cross section. If 
the spectrum of the accelerated
electrons is a relatively steep power law ($f(E) = \kappa E^{-\delta}, \delta 
\gg 1$),
then  most of the emission at photon
energy $k$ will come from electrons with energies $E$ just slightly above $k$.  
The bremsstrahlung  
emission can be then approximated as

\begin{equation}\label{asspec} 
	J(k) \simeq \kappa L
	\frac{16}{3} \alpha r_{0}^{2} n c \left\{ \begin{array}{ll}
	k^{-\delta-\frac{1}{2}}/(\sqrt{2}(\delta - \frac{1}{2})) & k \ll 1, \\ 		
	k^{-\delta}/ (\delta - 1) & k \gg 1. \end{array} \right.  
\end{equation} 
However, for $\delta < 1 $ most of the contribution at photon
energy $k$ will come from the highest energy electrons that this spectrum 
extends to. In our case this will be roughly the energy $E_{\rm synch}$ defined 
in previous section above which the electron spectra decrease rapidly. The 
photon spectrum in this case can be approximated as

\begin{equation}\label{asspec2} 
	J(k) \simeq \kappa L
	\frac{16}{3} \alpha r_{0}^{2} n c 
	\frac{E_{\rm synch}^{1-\delta}}{1-\delta} 
	k^{-1}\left\{ \begin{array}{ll}
	1 + \frac {(\delta - 
1)k^{-\delta+\frac{1}{2}}}{\sqrt{2}(\delta-\frac{1}{2})E_{\rm synch}^{1-\delta}} 
& k \ll 1, \\ 		
1 - (k/E_{\rm synch})^{1-\delta} & E_{\rm synch} \gg k \gg 1.\end{array} \right. 
\end{equation}
which for most practical purposes (for $k>0.03$ and $E_{\rm synch}>10^3$) gives 
$J(k) \propto k^{-1}$. 

Application of these equations to the emission from the acceleration site is 
straight forward because the spectrum of the accelerated electrons for most of 
the relevant energy range (specially those of interest here) can be approximated 
with simple broken power laws. Therefore, the emission $J_{AS}$ is obtained from 
the above expressions with the  replacing of $n$ by $n_{AS}$, the density at the 
acceleration site. For low values of  $\zeta_{\rm esc}$ we have steep electron 
spectra and a photon spectrum given by equation (\ref{asspec}). For high values 
the photon spectra are expected to obey 
equation (\ref{asspec2}). The dashed line in Figure 
6 shows the photon spectra of the acceleration site for the same 
model 
parameters as Figures 3 and 4 and demonstrate both of the above behaviors. 

The emission from the foot points is more complicated involving both $T_{\rm 
esc}$ and $\dot E_L$. If the  spectrum at the acceleration site, $f_{AS}(E)$, is 
relatively steep, which would be the case for $\zeta_{\rm esc} \ll 1$, 
and/or if $s+q' $ is small, then  the effective electron 
spectrum 
for {\bf the thick target } emission, equation (\ref{fthick}), can be 
approximated roughly as 
\begin{equation} \label{fthick2} 
f_{thick} \simeq 1/(-{\dot E}_L T_{\rm esc})\int_{E}^\infty f_{AS}(E')
dE' = - \kappa E^{-\delta + 1}/((\delta-1) {\dot E}_L T_{esc}(E)).
\end{equation}
Substitution of this in equation (\ref{FootPoint}), and again ignoring the 
variation of $T_{\rm esc}$ and the cross section, we find the following rough
approximation for the foot point emission at all $k$ and for $\delta > (5-s)/2$.

\begin{equation}\label{fpspec}
J_{FP} \simeq \kappa L \frac{16}{3} \frac{\alpha}{4 \pi \ln \Lambda} 
\frac{k^{-\delta 
+ 1}}{T_{\rm esc}(k)} \frac{1}{(\delta - 1)(\delta - 2)}.
\end{equation}

\begin{figure}[htbp]
\leavevmode
\centerline{
\psfig{file=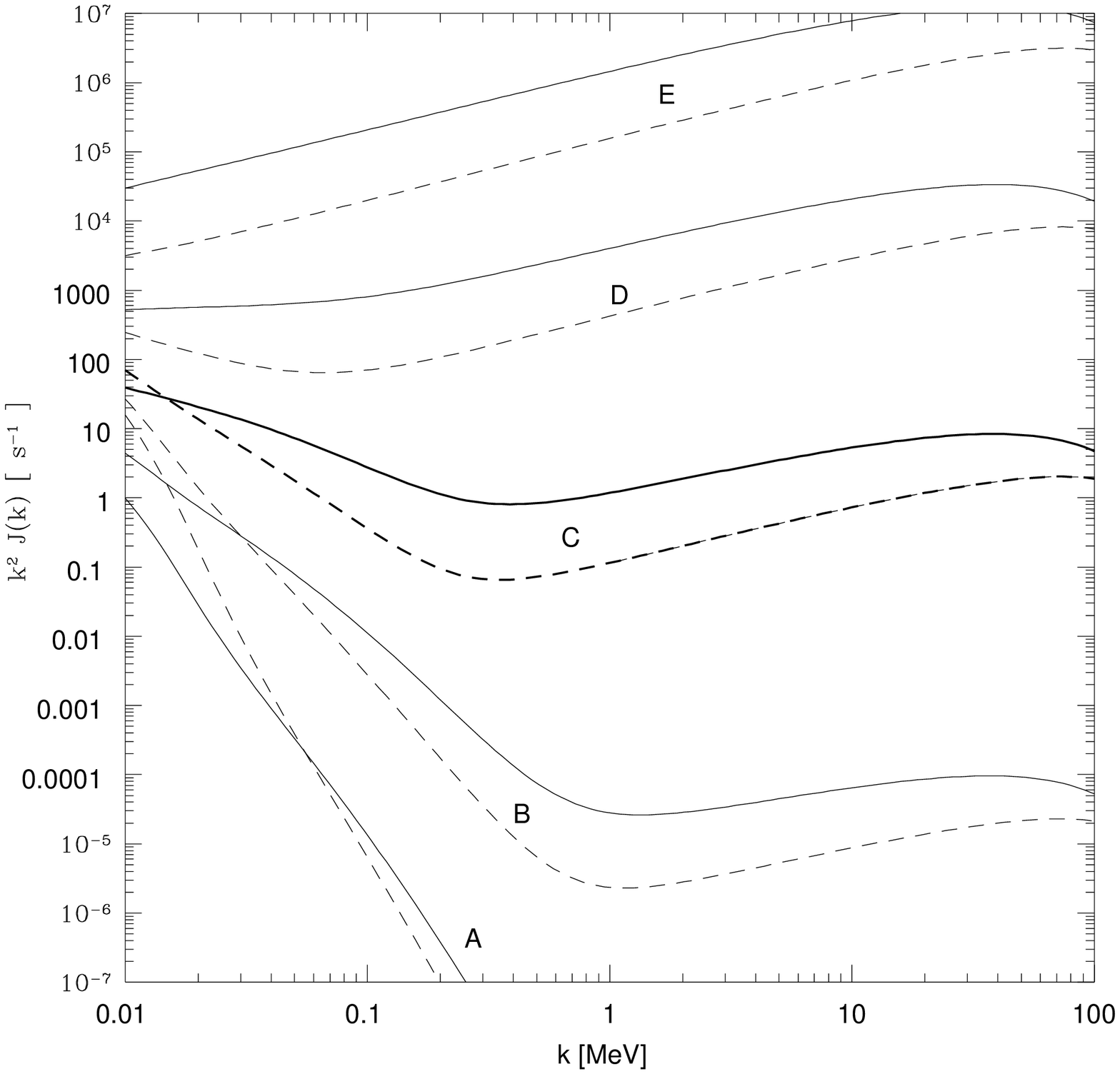,width=0.5\textwidth,height=0.5\textwidth}
\psfig{file=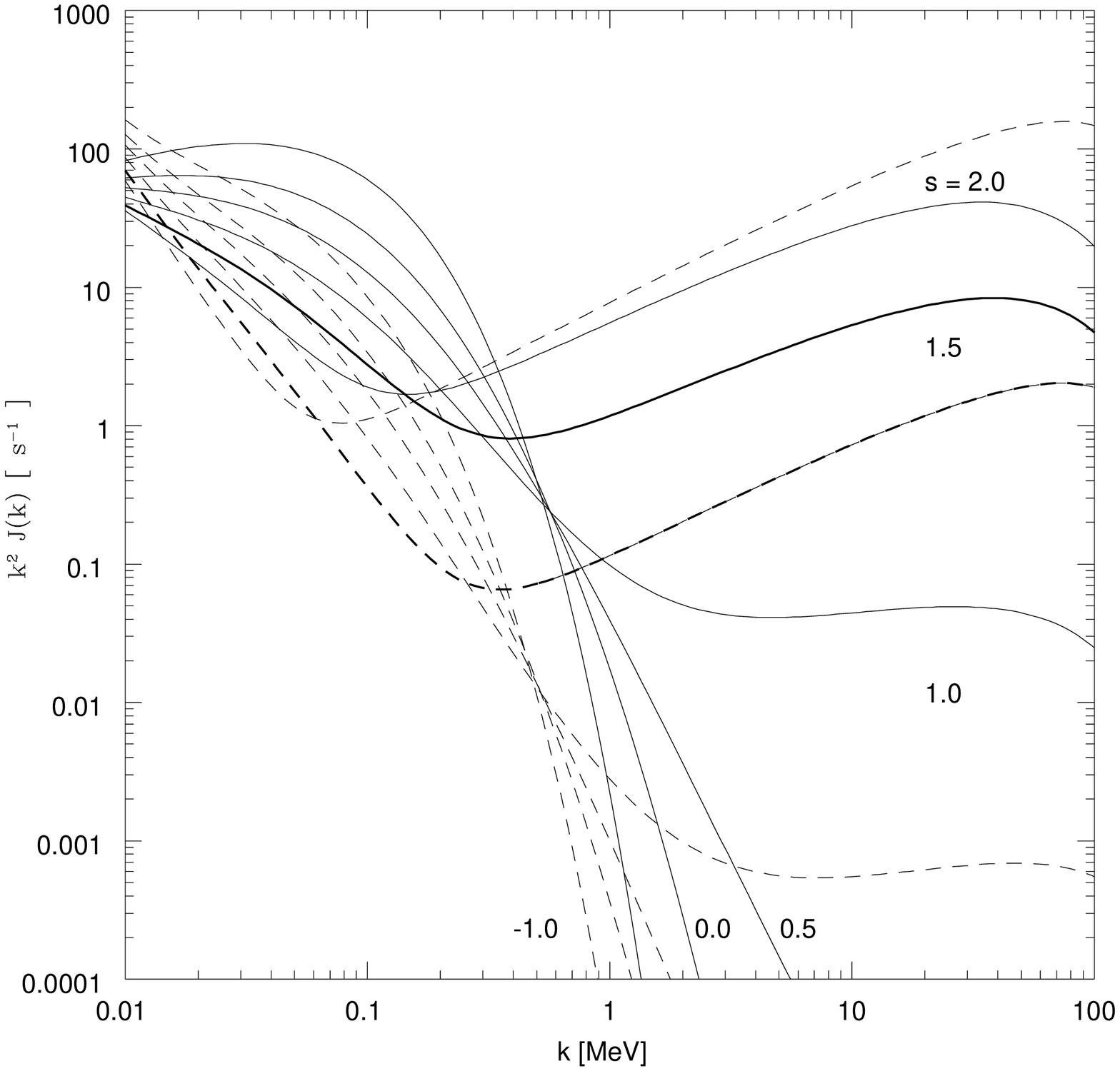,width=0.5\textwidth,height=0.5\textwidth}
} 
\caption{Thick target or foot point (solid lines) and thin target or 
acceleration site (dashed lines) photon spectra (multiplied by $k^2$, for 
convenience) for the  models of 
Figure \ref{particle} or the parameters given in Table 1. Note that for high 
values of $s$ and/or $\zeta_{\rm esc}$ both spectra vary as $J(k) \propto 
k^{-1}$ shown in equation (\ref{asspec2}).}
\label{spectra}
\end{figure}

However, as evident from Figures \ref{particle} and \ref{thick}  for high 
acceleration rates and/or 
slow escape $(R_{\rm esc} \gg 1)$ the accelerated electron spectrum flattens. In 
this case most of the contribution to the 
integrals for $f_{\rm thick}$  and for $J_{FP}$ (eqs. [\ref{fthick}] and
[\ref{FootPoint}], respectively) will come from the highest energy electrons and 
the photon 
spectrum, aside 
from the logarithmic and other slowly varying terms, will again be proportional 
to $k^{-1}$. These behaviors 
are seen in Figure \ref{spectra} where we plot the spectra of the foot point 
source (the 
solid lines) for the 
same 
parameters as the previous figures.

Using these simple approximations we can calculate the ratio of the  
foot point to loop top emissions. It is easy to show that in both 
nonrelativistic and 
extreme 
relativistic regions, and for steep spectra (small $\zeta_{\rm esc}$) this ratio 
(obtained from eqs. [\ref{fpspec}] and [\ref{asspec}]) can be given by the 
simple expression
\begin{equation}
\label{ratio}
\frac{J_{FP}}{J_{AS}} \simeq \frac{\tau_{\mathrm{Coul}} (k)}{T_{esc} (k)} = 
\frac{\zeta_{\rm Coul}}{\zeta_{\rm esc}}
\left\{ \begin{array}{ll} 2^{(2-s)/4}k^{2-s/2} & k \ll 1, 
\\k^{1-s} & k \gg 1,\end{array} \right.
\end{equation}
where we have neglected terms like $\delta -1$, $\delta -2$, etc. and have 
defined 
the Coulomb collision time scale
\begin{equation}\label{taucoul}
\tau_{\mathrm{Coul}} (k) \equiv -\left(\frac{E}{\dot{E} _{\rm Coul}} 
\right)_{E=k} = 
(4 
\pi r_{0}^{2} \ln \Lambda n_{AS} c)^{-1} \left\{ \begin{array}{ll}
			\sqrt{2} k^{3/2} & k \ll 1, \\
			k & k \gg 1.
		\end{array} \right.
\end{equation}
In addition to their simplicity, these expressions show that in case of 
relatively slow acceleration or rapid escape the ratio of the emission from the 
acceleration site to that of the foot 
points 
depends only on the product of the density and the escape rate ($n{\cal T}_{\rm 
esc} \propto \zeta_{\rm Coul}/\zeta{\rm esc}$) in the acceleration site and is
independent of the rest of the parameters of the loop or flare plasma.
This ratio is also independent of the acceleration rate $D(E)$ or the 
acceleration process as a whole. It is applicable to any 
acceleration model which produces a nearly isotropic distribution of electrons 
with a relatively steep spectrum.

\begin{figure}[htbp]
\leavevmode
\centerline{
\psfig{file=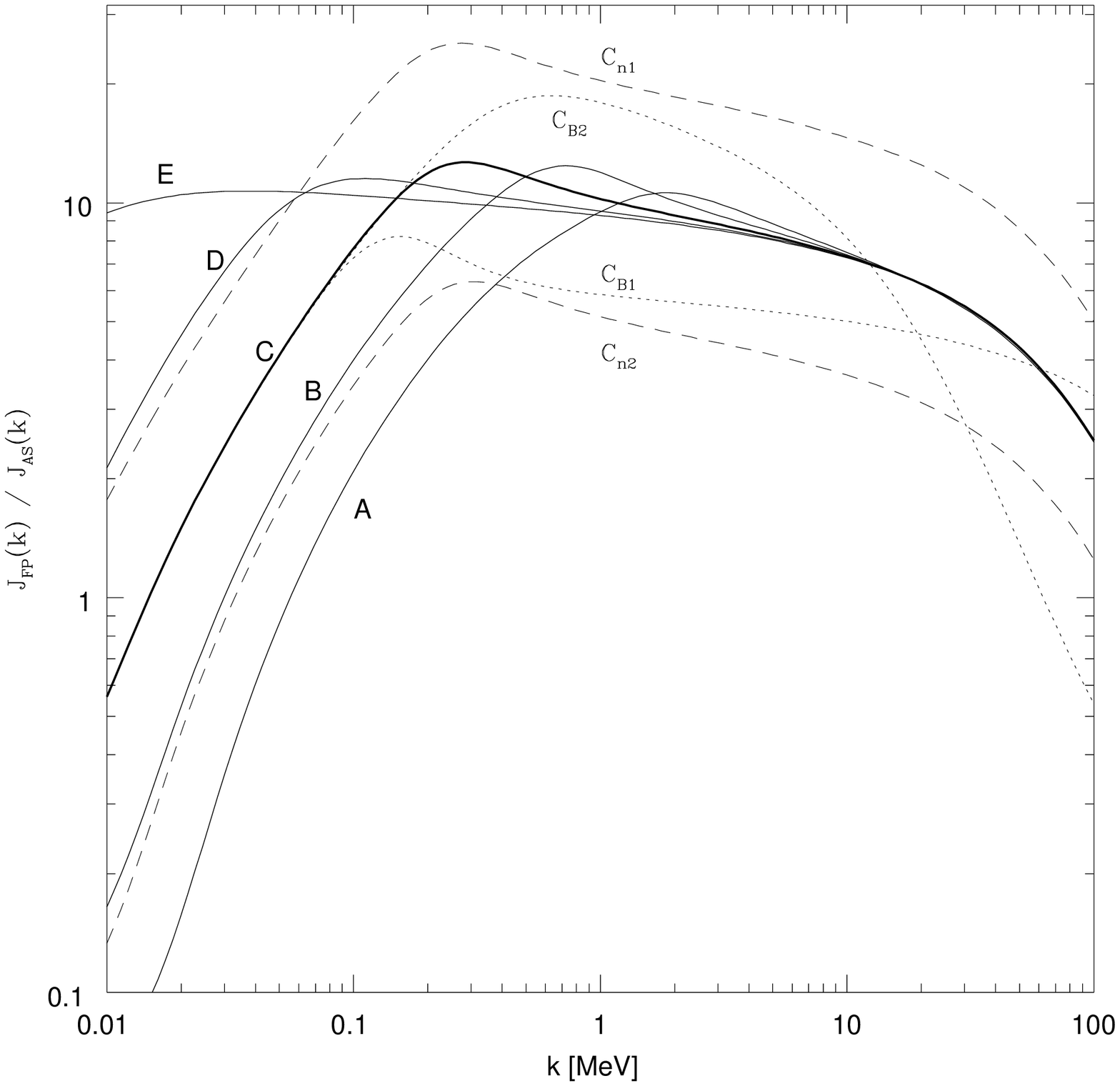,width=0.5\textwidth,height=0.5\textwidth}
\psfig{file=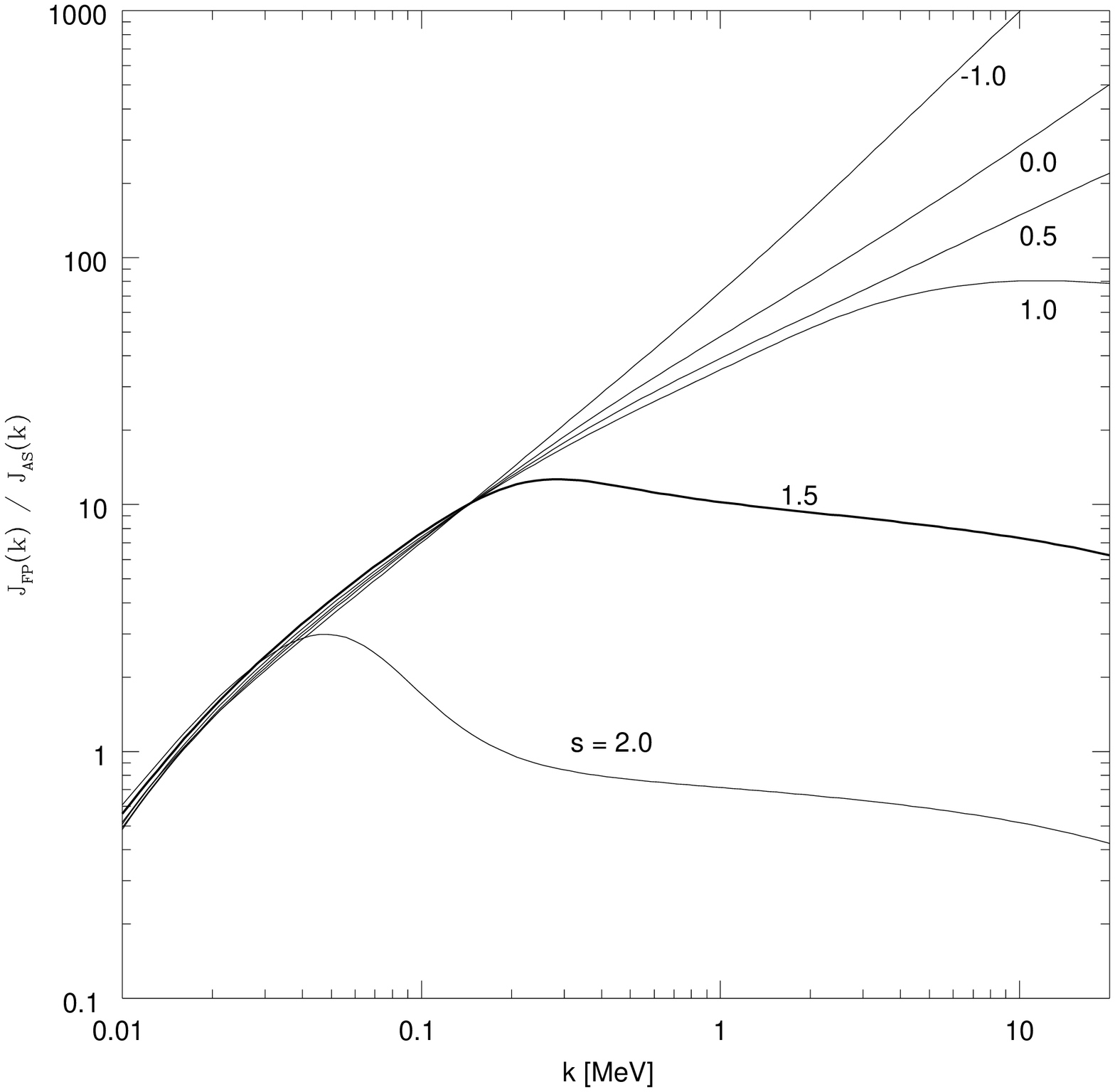,width=0.5\textwidth,height=0.5\textwidth}
}
\caption{The ratio of emission from the foot points to the loop top as a 
function of photon energy for the  models of 
Figure \ref{particle} or the parameters given  in Table 1. Note the presence of 
both rise-and-fall and nearly constant behaviors in the {\it left panel} as 
expected from 
equations (\ref{ratio}) and (\ref{ratio2}). Also note near independence from $s$ 
of the 
ratio at low energies in the {\it right panel} except for high values of $s$.}
\label{specratio}
\end{figure}

For larger values of $\zeta_{\rm esc}$ and flatter electron spectra both the
acceleration site and foot point spectra become proportional to $k^{-1}$ so that 
the ratio becomes independent of the photon energy (except at high energies 
where synchrotron 
losses dominate).
\begin{equation}
\label{ratio2}
\frac{J_{FP}}{J_{AS}} \propto \frac{\tau_{\mathrm{Coul}} (E_{\rm 
synch})}{T_{esc} (E_{\rm synch})} \propto   
\frac{\zeta_{\rm Coul}}{\zeta_{\rm esc}}
\end{equation}
Figure \ref{specratio} shows the variation of this ratio with the photon energy 
and other 
model 
parameters. Both a relatively constant ratio (expected for high $\zeta_{\rm 
esc}$, e.g. curve $E$), and rising-and-falling ratios similar to that expected 
from equation (15) (curves $A$ to $C$) are present in this figure. However, 
because in most cases the electron spectra ($f_{AS}, f_{\rm thick}$) suffer many 
deviations from  pure steep power laws, the numerical results deviate 
significantly from the simple analytic expressions (except, perhaps at extreme 
high and low values of $\zeta_{\rm esc}$) and show sensitivity to other 
parameters. This fact and  the diversity of the shapes shows that,
in general, the relative strengths and spectral shapes of the foot point and 
loop top emissions can be good diagnostic tools for the investigation 
of the acceleration, escape and energy loss  processes.

\section{Comparison With YOHKOH Data}

In this section we compare some of the model results with the characteristics 
of the limb flares observed by Masuda (1994). As mentioned in \S 1, six out of 
ten flares studied showed detectable emission from top of the loop as well as 
the usual foot point emission. We use the ratio of foot point to loop top 
emission in the the energy bands of YOHKOH and the spectral indices whenever 
these are available. 
\begin{figure}[htbp]
\leavevmode
\centerline{
\psfig{file=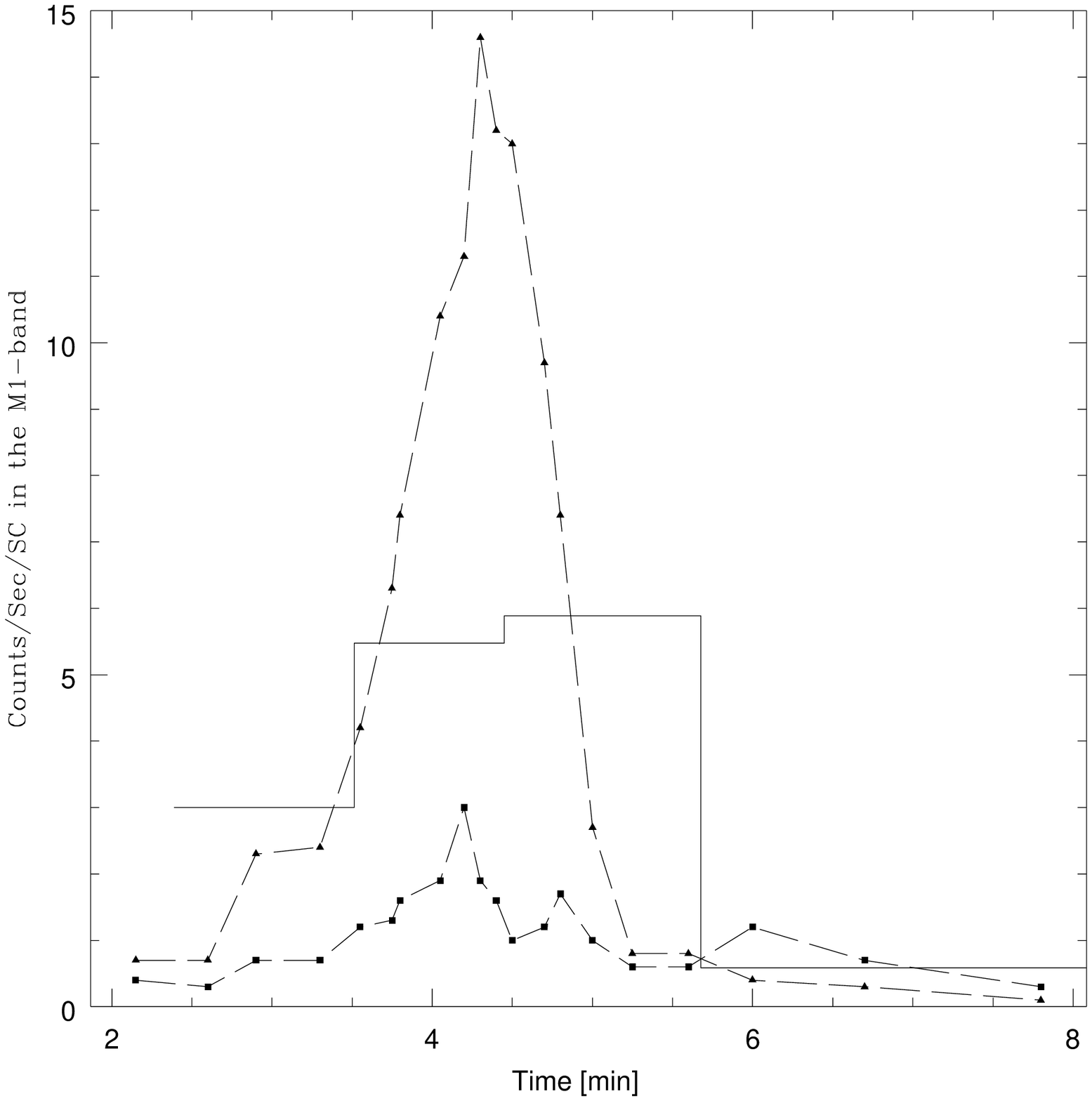,width=0.5\textwidth,height=0.5\textwidth}
\psfig{file=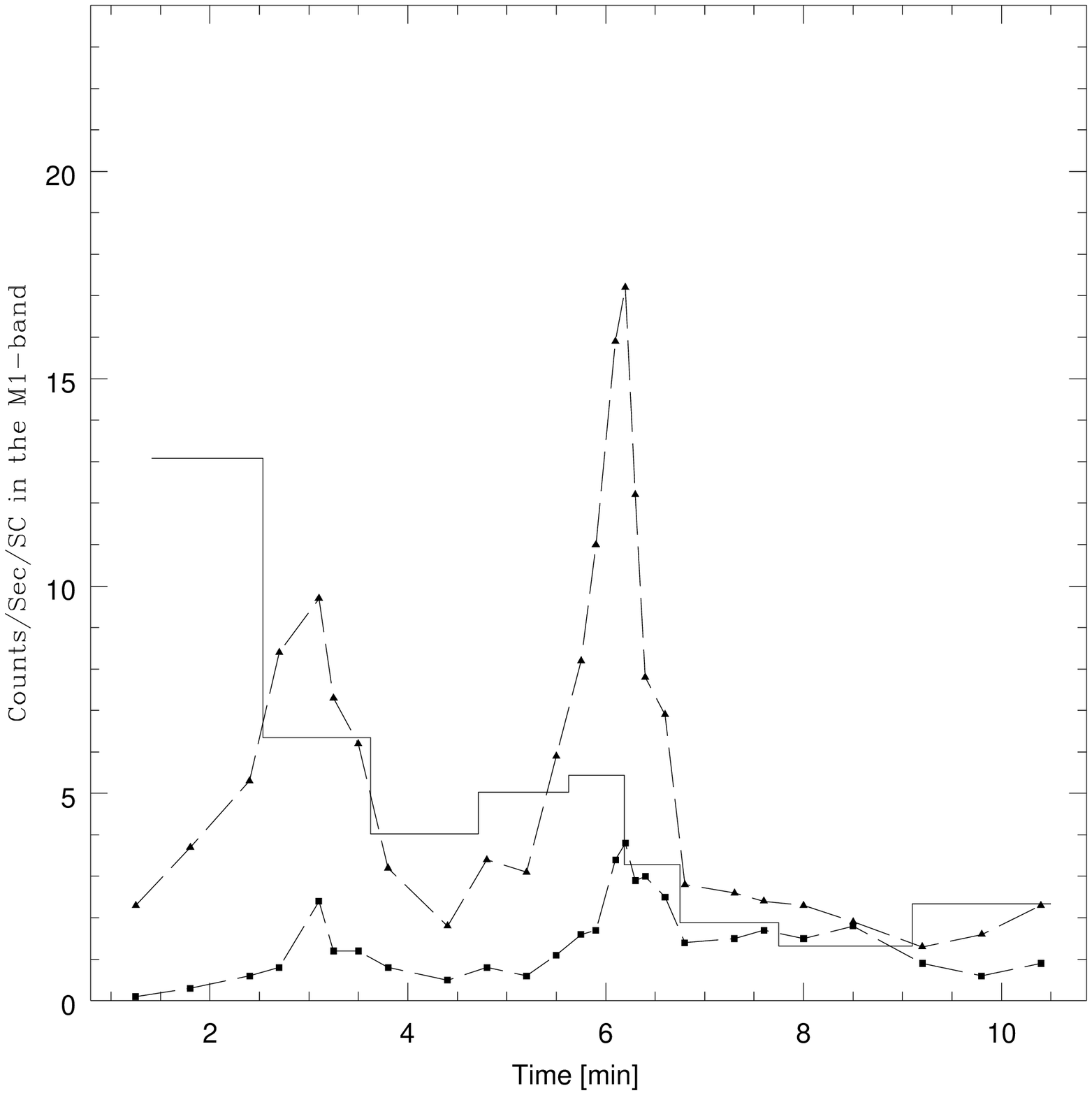,width=0.5\textwidth,height=0.5\textwidth}
}
\caption{The time profiles of two flares as seen by YOHKOH in the M1 band (23-33 
keV). 
The triangles represent the foot point flux and the squares represent the loop 
top flux.
The  solid histogram gives the ratio of the foot point to the loop top 
fluxes: {\it Left Panel:} January 13, 1992; {\it Right Panel:} October 4, 1922. 
From Masuda (1994).}
\label{lightcurves}
\end{figure}

Figures \ref{lightcurves} show the observed variation with time of 
the above mentioned ratio for 
channel M1 (23-33 keV) in two flares taken from Masuda (1994). As evident when 
there is 
sufficient signal from both sources the ratio is always greater than one and 
averages around 5. Alexander \& Metcalf (1997) have reanalized the January 13, 
1992 flare using a different image reconstruction technique. They obtain 
slightly different light curves and a variable ratio in the range 3 to 4.
Of course there is a strong  observational selection bias 
here; the ratio is much larger than this for flares with undetectable 
emission from the loop top. There should, however, be no strong bias against 
flares with values of this ratio less than one. In general, as the flare 
progresses the ratio tends 
to decrease (and becomes equal or less than one specially in the L channel; 
14-23 keV). 
Most of this trend is probably due to increasing thermal emission from the loop 
tops by the evaporated plasma. But some of it could be due to changes in the 
acceleration or other flare parameters.

Masuda also calculates average photon spectral indices by fitting a 
power law to the photon number fluxes in two adjacent channels. In Table 2 we 
present some 
of these values for channels M1 and M2 from the foot point and loop top sources. 
In general the spectra appear to be steep. Some of this could 
be due to contamination by thermal plasma especially for flares where the 
impulsive phase is not well isolated from the thermal phase. We will ignore 
values greater than 6. Furthermore, 
because of the low counts the error bars on these values (not given in Masuda 
1994) must be considerable. Alexander \& Metcalf (1997) carry out a more detail 
analysis of the spectra of January 13, 1992 flare. They give directly the values 
of the power law index of the injected electrons in generic thin and thick 
target models. Using equation (\ref{LoopTop}) and ignoring the logarithmic 
dependence in equation (\ref{crosssec}) it is easy to show that for electron 
spectrum $f\propto E^{-\delta}$, the thin target photon spectrum will be 
$J\propto k^{-\delta - 1/2}$. For the peak of the emission in channel M1 
Alexander \& Metcalf give $2.5 < \delta_{LT} < 3.7$ and $3.2 < \delta_{FP} < 
4.4$, which translate to average photon indices of 3.6 and 4.3 for loop top and 
foot point sources, respectively. Considering the error bars, this is in good 
agreement with the Masuda results.

\begin{center}
\centerline{TABLE 2}
\centerline{Photon spectral indices for selected flares from Masuda (1994)}
\begin{tabular}{|l|c|c|} \hline\hline
Date & FootPoint & LoopTop \\ \hline
91/12/02 & 6.3  & 5.5    \\
91/12/15 & 3.9  & -      \\
92/01/13 & 4.0  & 4.1    \\
92/04/01 & 3.7  & -      \\
92/10/04 & 3.4  & 5.2    \\
92/11/23 & 4.6  & 6.3    \\ \hline\hline
\end{tabular}
\end{center}

\begin{figure}[htbp]
\leavevmode\centering
\psfig{file=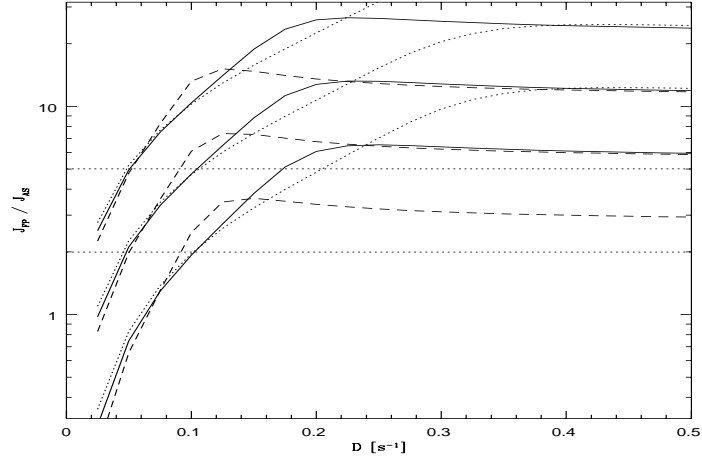,width=0.6\textwidth,height=0.4\textwidth}
\caption{The ratio $J_{FP} / J_{LT}$ in the 23 to 33 keV range (the M1 band) as 
a function of $ {\cal D}$ for three  
escape times ${\cal T}_{\rm esc}= 0.05, 0.1$  and 0.2 s; dotted, solid and 
dashed lines, respectively, and for three values of density $n/(10^{10} {\rm 
cm}^{-3})=2.5, 5$ and 10; in each case from top to bottom, respectively.  The 
magnetic field $B=300$G, $s=1.5$ and $q' =1.7$. The horizontal dotted lines 
show the  range of the observed ratios.
}
\label{realratio}
\end{figure}

In what follows we try to get a rough measure of model parameters using 
the overall ranges of the intensity ratios and spectral indices. Figure 
\ref{realratio} shows 
the  intensity ratio, $J_{FP}/J_{LT}$, in the M1 band as a function of 
acceleration rate parameter 
${\cal D}$  for 
three different values of the escape time parameter ${\cal T}_{\rm esc}$ and 
density.
For small values of ${\cal D}$ or $\zeta_{\rm esc}$, i.e. for slow acceleration 
rates, this ratio varies rapidly and is sensitive to density and not the escape 
time ${\cal T}_{\rm esc}$. For  ${\cal D}>0.1$ (i.e. larger $\zeta_{\rm esc}$) 
this ratio approaches an asymptote whose value depends linearly on the ratio 
$\zeta_{\rm Coul}/\zeta_{\rm esc} \propto (n{\cal T}_{\rm esc})^{-1}$, in 
agreement with equation (\ref{ratio2}). Ratios of 1 
to 6 can be obtained for both low and high values of density and ${\cal D}$ (or 
$\zeta_{\rm esc}$). In the first case from this ratio one can determine the 
parameters $n$ and  ${\cal D}$ and in the latter case only the product $n{\cal 
T}_{\rm esc}$. 

\begin{figure}[htbp]
\leavevmode
\centerline{
\psfig{file=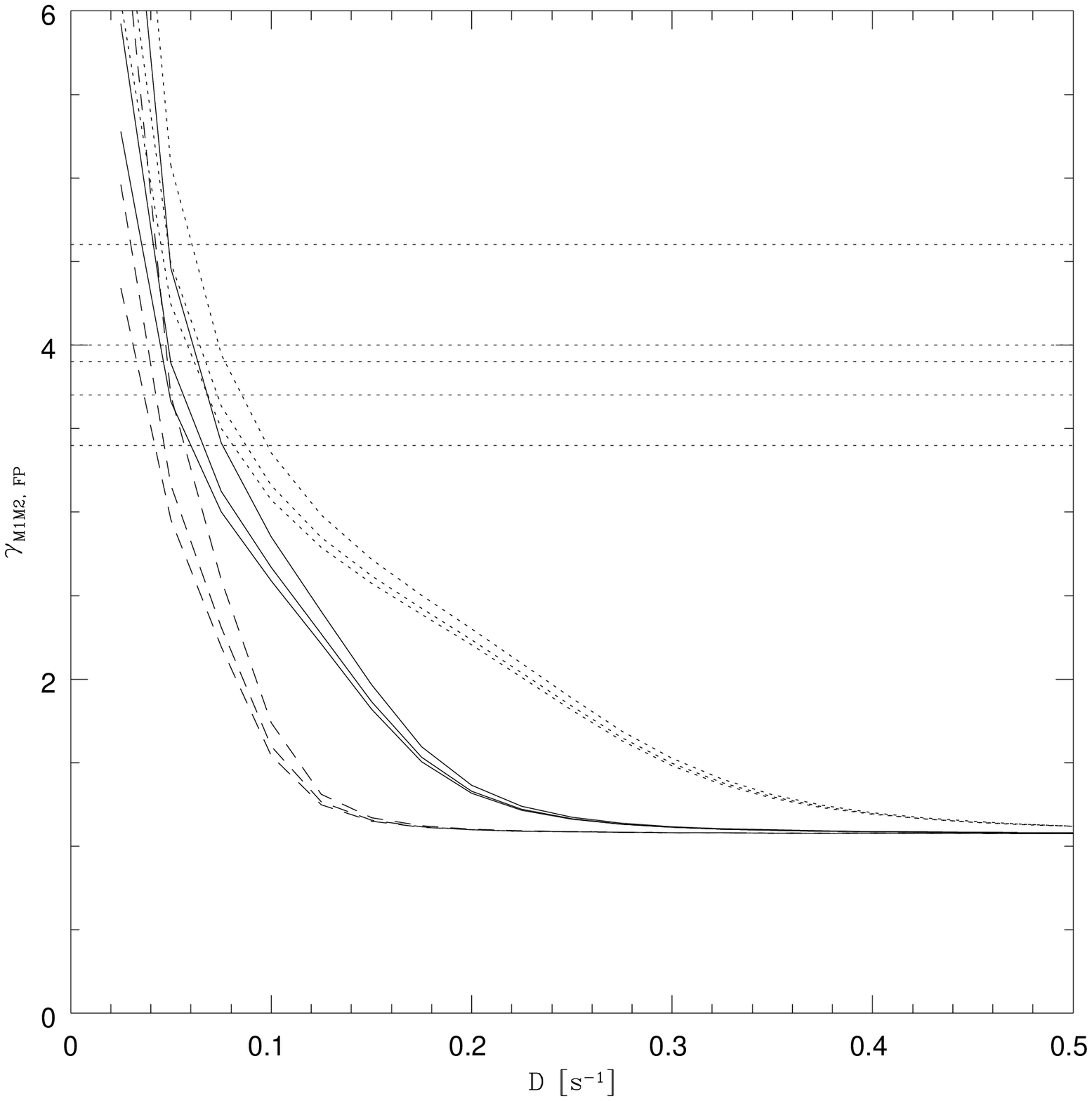,width=0.5\textwidth,height=0.5\textwidth}
\psfig{file=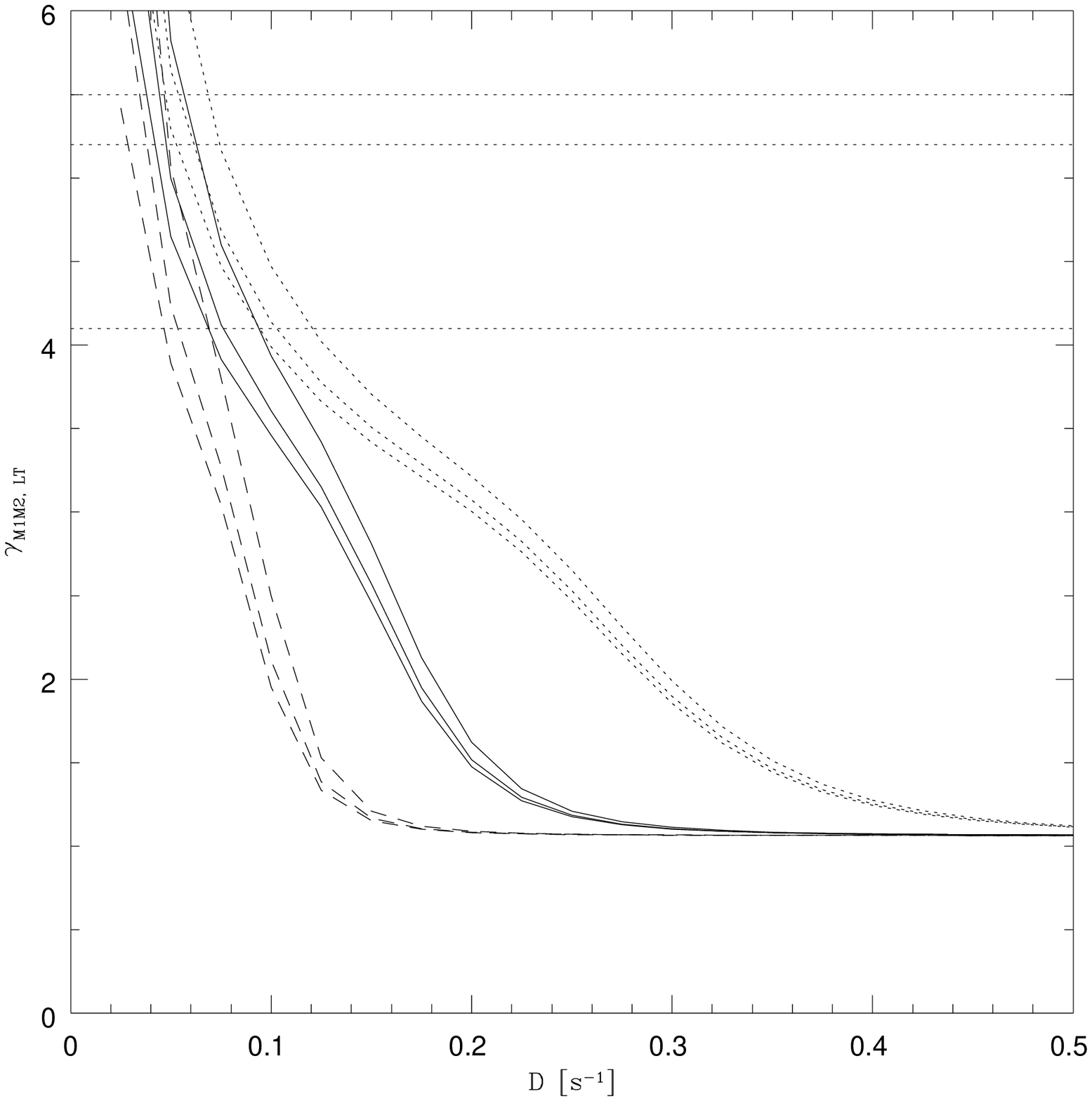,width=0.5\textwidth,height=0.5\textwidth}
}
\caption{Same as Figure \ref{realratio} but for the photon number spectral index 
 ($\gamma =- d{\rm ln}J(k)/d{\rm ln}k$) between channels M1 and M2. Note that 
the order of the densities here is the opposite of that in Figure 
\ref{realratio}, increasing from bottom to top. The horizontal dotted lines show 
the observed 
values of this index given in Table 2. {\it Left Panel:} Foot point source {\it 
Right Panel:} Loop top source.}
\label{realindex}
\end{figure}

This degeneracy can be 
broken from consideration of the spectral indices. Figures \ref{realindex} show 
the 
variations
of the photon number (not energy) spectral indices obtained from channels M1 and 
M2 (33-53 keV) for the foot 
point (keft panel) and loop top (right panel)sources. As in Figure 
\ref{realratio} the spectral index is 
plotted as 
a function of ${\cal D}$ for three values of ${\cal T}_{\rm esc}$ (dashed, solid 
and dotted lines) and density $n$.
The dotted horizontal lines show the observed values given in Table 2.
The curves here decrease monotonicly with ${\cal D}$  and approach the 
asymptotic value of $\gamma = -d{\rm ln}J/d{\rm ln}k \rightarrow 1$ at high 
${\cal D}$ values, as expected from equation 
(\ref{asspec2}). The observations on the other hand favor values of ${\cal D}$ 
 in the lower range (${\cal D} < 0.15$ s$^{-1}$) and values for other parameters 
which are similar to the values obtained from 
the spectral fittings by PPS. These values are well within the acceptable 
ranges. For example, the 
January 13, 1992 flare with the ratio of 5 and spectral indices of 4.1 and 4.0 
for 
the loop top and foot point sources, respectively,  can be obtained with a 
model with ${\cal D} \simeq 0.08{\rm s}^{-1}, {\cal T}_{\rm esc} \simeq 0.05{\rm 
s}, s=1.5, 
q' =1.7$ 
and $n \simeq 3\times 10^{10} {\rm cm}^{-3}$, and for wide ranges for $B$ and 
$L$.
Similarly the 
October 4, 1992 flare with the ratio of about 4 and spectral indices of 5.2 and 
3.4, respectively,  can be fitted  with almost same
model parameters except ${\cal D} \simeq 0.09{\rm s}^{-1}, {\cal T}_{\rm esc} 
\simeq 0.07{\rm s}$ and $n \geq 5\times 
10^{10} {\rm cm}^{-3}$. These values of ${\cal D}$, and the field strength of 
about 300G, require a turbulence energy density of $<10^{-5}(B^2/8\pi)$ (cf, 
PPS).

\begin{figure}[htbp]
\leavevmode
\centerline{
\psfig{file=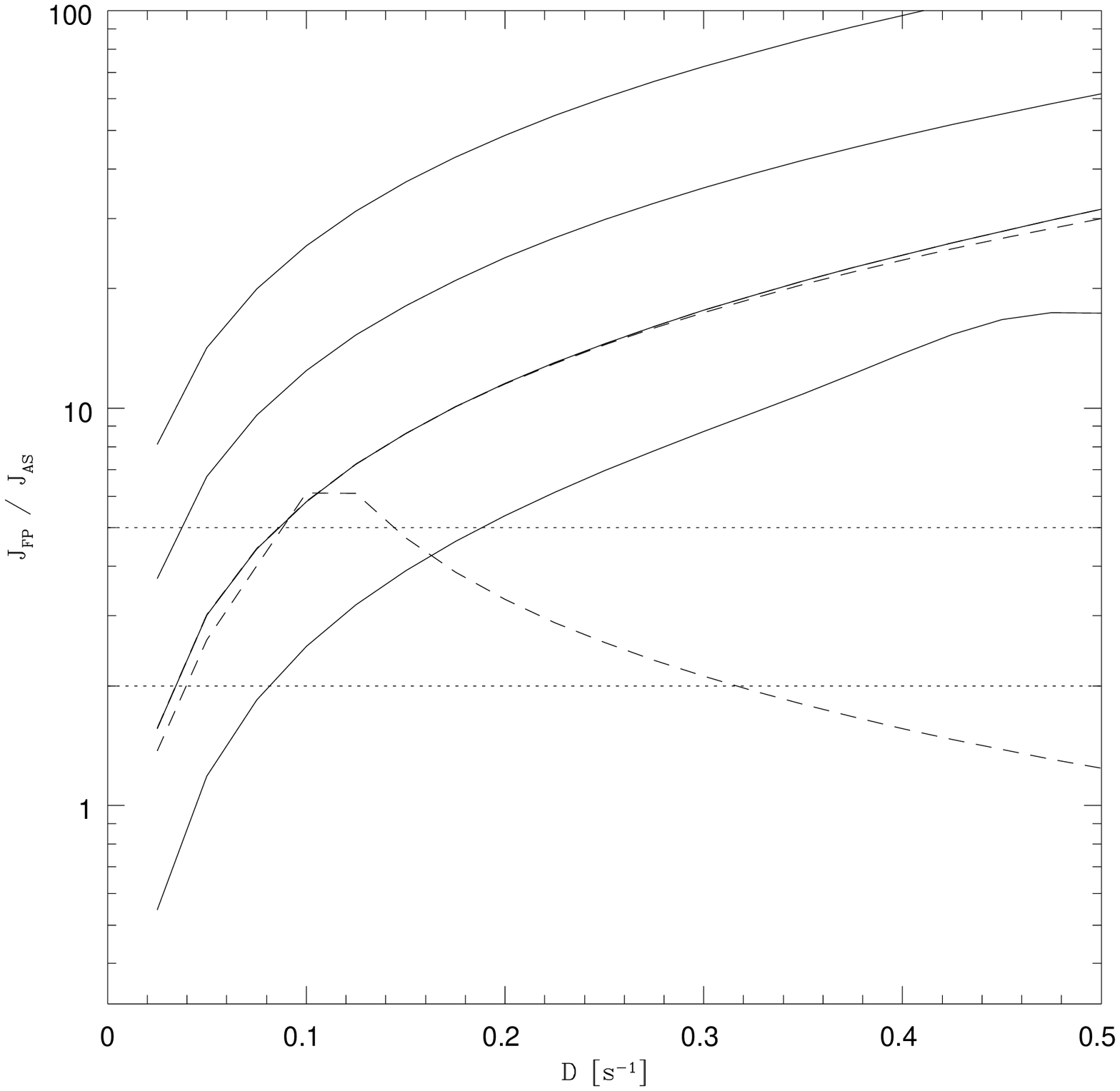,width=0.5\textwidth,height=0.5\textwidth}
\psfig{file=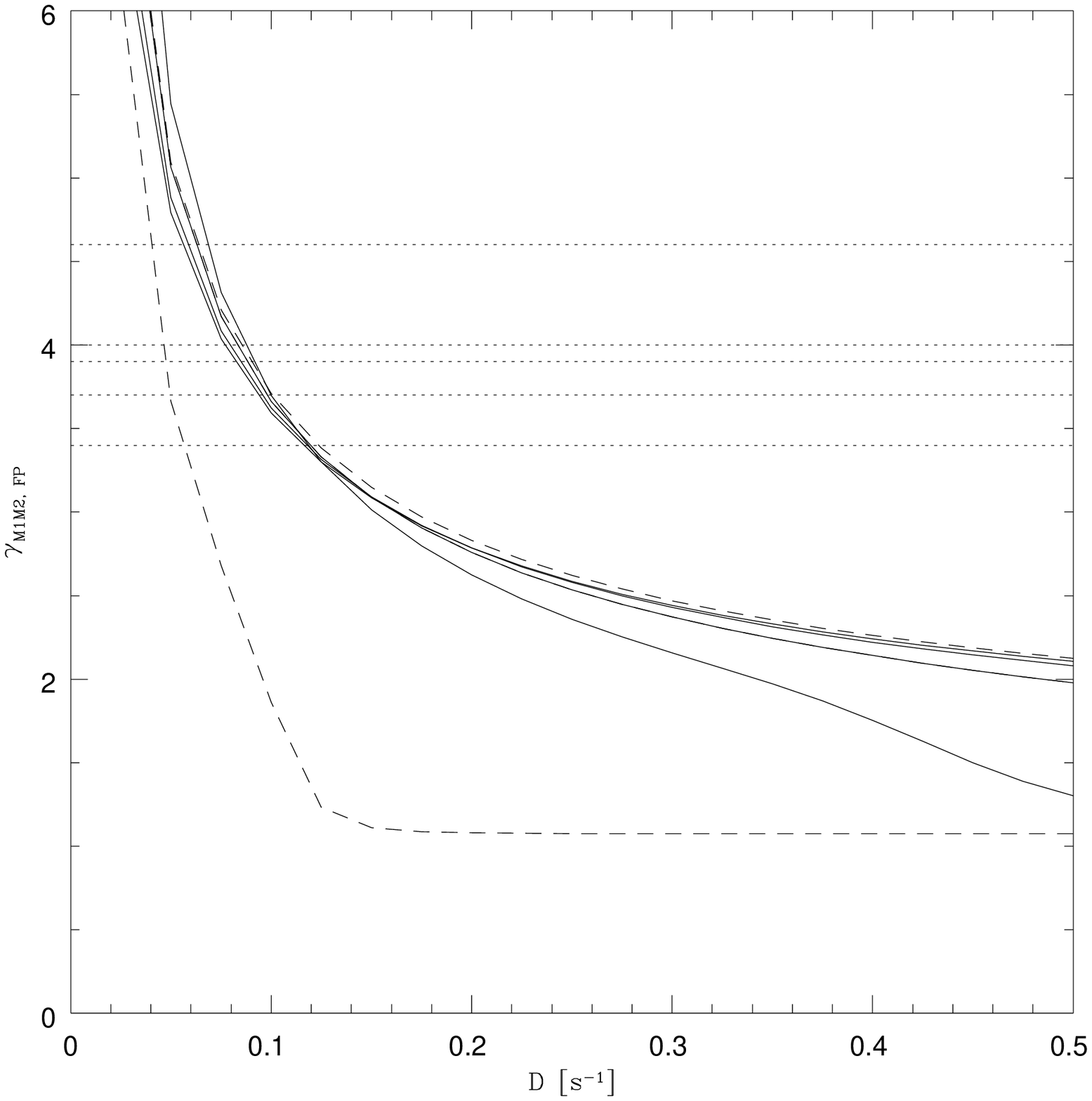,width=0.5\textwidth,height=0.5\textwidth}
}
\caption{{\it Left Panel:} Same as Figure \ref{realratio}, except using the 
whistler wave 
acceleration model. The solid lines from top to bottom are for densities 
$n/(10^{10} {\rm cm}^{-3})=1.25, 2.5, 5$ and 10, respectively. $B=300$G, 
$L=10^9$ cm and $s=q' =1.5$. The dashed lines show the effect of the 
magnetic field; $B=1000$G, upper curve; $B=100$G, lower curve.
{\it Right Panel:}Variations of the photon number spectral 
index of the 
foot point source between channels M1 and M2 for teh same model parameters as 
the left panel. Note that the increasing order of 
the density is for the high {\cal D} end of the solid lines.}
\label{whistler}
\end{figure}

We repeat this analysis in Figure \ref{whistler} for the whistler model. A wider 
range of behavior is seen here because of the dependence of the ${\cal T}_{\rm 
esc}$ on the magnetic field, size and density (see, e.g. eq. [10] in PPS). 
Similar parameter values are obtained for the observed flares except in this 
case the magnetic field can be constrained to be greater than 100G.

\section{Summary and Discussion}

In this paper we investigate the spatial structure of the impulsive phase hard 
X-ray emission. This work is motivated by the YOHKOH observations of hard X-rays 
from top of flaring loops, presumably where particle acceleration takes place, 
as 
well as from their foot points, as is seen more commonly. 

We first discuss the effects of particle transport in the usual thick target 
model whereby electrons are injected at the top of a flaring loop. We summarize 
the past results mainly from unpublished parts of J. Leach's (1984) thesis 
showing that 
a significant loop top emission, without any detectable emission from the legs 
of the loop, is possible in inhomogeneous loops where there is either a 
significant field convergence from the top to foot points or there is a strong 
density enhancement at the top of the loop. The former is a more natural 
expectation than the latter.

We point out two other possible explanations. The first is due to plasma 
turbulence 
at the acceleration site which can scatter and temporarily trap the electrons 
near the top of the loop giving rise to an enhanced emission as observed by 
YOHKOH. The second is if the accelerated particles have pancake type pitch angle 
distribution perpendicular to the field lines. Both of these could be a natural 
consequence of the stochastic acceleration model that has been investigated by 
Pertosian and colleagues (see e.g. Petrosian 1996 or PPS and references cited 
there) and Miller and colleagues (Miller et al. 1996 and Miller \& Reames 1996). 

A pancake type distribution can come about if acceleration is 
due to resonant scattering of electrons by plasma waves propagating 
perpendicular to the magnetic fields. As shown recently by Pryadko \& Petrosian 
(1999) such interactions preferentially accelerate electrons with  pitch  
angles near $90^\circ$. In this paper we have considered the more general case 
of 
stochastic acceleration that leads to an isotropic  pitch angle distribution, in 
which case the 
loop top or acceleration site emission is a direct consequence of the  
scattering effects of the acceleration agent.

Following the formalism developed in PPS we evaluate separately the spectrum of 
the radiation expected from the acceleration site and the foot points. We 
evaluate the relative strengths and spectral shapes of these two emissions and 
the dependence of these on the parameters describing the acceleration process 
and the flare plasma. The most important parameter is the product of the 
acceleration rate and the escape time; its value and energy dependence. We have 
adopted general parametric expressions for these quantities. In reality, as 
described by Pryadko \& Petrosian (1997, 1998 and 1999) the determination of 
these parameters, which are obtained from the Fokker-Planck coefficients, is 
complicated and requires extensive numerical analysis. They depend strongly on 
the ratio of the plasma to gyro-frequency, or $n/B^2$. The plasma density  $n$ 
and the field strength $B$ also play important roles through their influence on 
the losses due to Coulomb collisions and synchrortron emission. The parametric 
forms chosen here are designed to mimic the behavior of the more accurate 
numerical results.

We first give a general description of the characteristics of these emissions 
and 
compare them with some approximate analytic expressions. We then compare the 
theoretical ratio of the intensities and spectral indices of the X-ray emissions
with YOHKOH observations. We demonstrate how the above mentioned parameters can 
be constrained with such observations. The ranges of the parameters favored here
is similar to those obtained through spectral fittings by PPS. The current 
spatially resolved information is not sufficient for setting strong constraints 
on the acceleration or flare plasma parameters or for distinguishing the 
stochastic acceleration model in general from other acceleration and transport 
models. Our purpose here has been to demonstrate the feasibility of this model 
and point out how future combined high spatial and spectral resolution data, for 
example those expected from HESSI, can answer these important questions.

Aside from using the simple parametric approximations mentioned above we have 
also used the assumption of pitch angle isotropy. At low energies this 
assumption may break down. As shown by Pryadko \& Petrosian (1997) for waves 
propagating along the magnetic field lines the acceleration time scale may 
become smaller than the scattering time and the latter may become longer than 
the traverse time $\tau_{tr}\sim L/v$. To account for this we have added the 
additional term to the escape time which is proportional to $\tau_{tr}$. This 
does not fully take care of the problem arising with the possible anisotropy of 
the pitch angle distribution. Anisotropy can also arise in the presence of waves 
propagating perpendicular to the field lines because such waves accelerate 
electrons preferentially with pitch angles $\alpha \sim 90^\circ$ (Pryadko \& 
Petrosian 1999). As mentioned above this can give rise to pancake type 
distribution which also can 
give rise to an enhanced emission near the acceleration site. The effects of 
these complications will be addressed in future publications.

This work is supported in parts by NASA grant NAG-5-7144-0002, Lockheed grant  
SA30G730R-AO7, and grants from 
Stanford's Undergraduate Research Opportunities and Bing Summer Science 
Fellowship Programs. TQD would like to thank Brian Park and Nicole Lloyd for 
helpful discussions on computational aspects of this work.

\end{document}